\documentclass[twocolumn]{aastex62}
\usepackage{multirow}
\usepackage{soul}
\usepackage{appendix}

\shortauthors{Tremblay et al.}

\begin{document}

\title{\textbf{The Detectability and Constraints of Biosignature Gases \\ in the Near \& Mid-Infrared from Transit Transmission Spectroscopy}}

\author[0000-0002-1715-1986]{L. Tremblay}
\affil{School of Earth and Space Exploration, Arizona State University, \\ PO Box 871404, Tempe, AZ 85287-1404, USA}
\author{M.R. Line}
\affil{School of Earth and Space Exploration, Arizona State University, \\ PO Box 871404, Tempe, AZ 85287-1404, USA}
\author{K. Stevenson}
\affil{Space Telescope Science Institute, \\ 3700 San Martin Drive, Baltimore, MD 21218, USA}
\author{T. Kataria}
\affil{Jet Propulsion Laboratory, California Institute of Technology, \\ 4800 Oak Grove Drive, Pasadena, CA 91109, USA}
\author{R.T. Zellem}
\affil{Jet Propulsion Laboratory, California Institute of Technology, \\ 4800 Oak Grove Drive, Pasadena, CA 91109, USA}
\author{J.J. Fortney}
\affil{Department of Astronomy \& Astrophysics, University of California Santa Cruz, \\ Santa Cruz, CA, USA}
\author{C. Morley}
\affil{Department of Astronomy, University of Texas at Austin, \\ Austin, TX, USA}

\begin{abstract}

The James Webb Space Telescope (\textit{JWST}) is expected to revolutionize our understanding of Jovian worlds over the coming decade. However, as we push towards characterizing cooler, smaller, ``terrestrial-like" planets, dedicated next-generation facilities will be required to tease out the small spectral signatures indicative of biological activity. Here, we evaluate the feasibility of determining atmospheric properties, from near-to-mid-infrared transmission spectra, of transiting temperate terrestrial M-dwarf companions. Specifically, we utilize atmospheric retrievals to explore the trade space between spectral resolution, wavelength coverage, and signal-to-noise on our ability to both detect molecular species and constrain their abundances. We find that increasing spectral resolution beyond R=100 for near-infrared wavelengths, shorter than 5$\mu$m, proves to reduce the degeneracy between spectral features of different molecules and thus greatly benefits the abundance constraints. However, this benefit is greatly diminished beyond 5$\mu$m as any overlap between broad features in the mid-infrared does not deconvolve with higher resolutions. Additionally, our findings revealed that the inclusion of features beyond 11$\mu$m did not meaningfully improve the detection significance nor abundance constraints results. We conclude that an instrument with continuous wavelength coverage from $\sim$2-11$\mu$m, spectral resolution of R$\simeq$50-300, and a 25m$^2$ collecting area, would be capable of detecting H$_2$O, CO$_2$, CH$_4$, O$_3$, and N$_2$O in the atmosphere of an Earth-analog transiting a M-dwarf (mag$_{K}=8.0$) within 50 transits, and obtain better than an order-of-magnitude constraint on each of their abundances. 

\end{abstract}


\section{Introduction}

Characterizing the climate and composition of terrestrial atmospheres is a primary goal of exoplanet science over the next decades (NAS 2020 Vision\footnote{\url{https://sites.nationalacademies.org/cs/groups/bpasite/documents/webpage/bpa_064932.pdf}}). \textit{Temperate} terrestrial exoplanets are of course, of great interest due to their possible astrobiological implications. Namely, their potential to develop and maintain volatile-rich secondary atmospheres capable of supporting biology. As Earth is the only known planet to host life, it is straight-forward that we would begin our search for extraterrestrial life by considering the conditions on Earth as the primary conditions which must exist for the development of life as we know it.

The principal gases that comprise Earth's atmosphere (nitrogen/nitrous oxide, water, carbon dioxide, oxygen/ozone, and methane) are generally regarded as ``bio-indicators" meaning that their presence is indicative of potentially habitable atmospheric conditions. ``Biosignatures" are those observables that are representative of biological processes affecting the planet's atmospheric chemistry (i.e. the presence of life). \cite{CatKast07} showed that all of the bulk gases in Earth's atmosphere, excluding inert gases, are influenced by biogenic processes. Life on Earth is the primary contributor to the global chemical disequilibrium \citep{krissTott16}, and for this reason disequilibrium is presumed to be a robust biosignature \citep{Cockell09, Kasting09,Leger00,Sagan93,Seager14, Seager15,Seager10}. The combinations of molecules responsible for Earth's disequilibrium are O$_{3}$+CH$_{4}$ and O$_{3}$+N$_{2}$O \citep{krissTott16}. If detected at certain abundances, the presence of these molecules simultaneously cannot be plausibly explained by abiotic processes  \citep{DomGold14,Harman15,Harman18,Leger93,Meadows18,Meadows17,Tian14,Schwiterman16}.  As such, a key goal of exoplanet exploration is to quantify the presence of these gases on nearby terrestrial worlds.

Due to the favorable planet-to-star radius ratio and occurrence rates, M-dwarf systems offer the best near-term opportunity for characterizing temperate terrestrial worlds and the possible presence of biosignatures.  Ground-based radial velocity surveys and \textit{TESS} are expected to detect dozens of temperate terrestrial-sized (0.8 - 1.6R$_{\Earth}$) exoplanets around M-dwarf's \citep{Quirrenbach18,Barcklay18,Kempton18,Zechmeister19}, potentially amenable to follow-up with \textit{JWST}. Numerous previous works have demonstrated that \textit{JWST} will likely provide astounding constraints on the atmospheric properties of Jovian-to-Super-Earth planets  \citep{Deming09,Beichman14,Barstow14,Greene16,Batalha18,Rochetto16,Benneke12}. However, it remains an open question as to how well \textit{JWST} will perform as it is pushed to observe temperate terrestrial bodies with smaller spectral features. Possible limitations may stem from a combination of large detector noise floors (similar detector technology as \textit{HST WFC3} and \textit{Spitzer IRAC}, \cite{Greene16}), saturation limits for the brightest targets \citep{Batalha18}, and lack of continuous near-to-mid-infrared wavelength coverage in a single mode (greatly increasing the number of required transits).

The discovery of the TRAPPIST-1 system \citep{Gillon16} has lead to a number of works investigating the climate, composition, and internal structure of temperate M-dwarf worlds \citep{Unterborn17,Turbet18,Suissa18b,Meadows17,Wunderlich19}. Complementary to these efforts, recent studies have explored the capabilities of \textit{JWST} to characterize various atmospheric compositions on TRAPPIST-1 planets through both transmission and emission spectroscopy \citep{Barstow16,Morley17,Batalha18,KT18,LustYeag19,Wunderlich19}.

\cite{Morley17} concluded that \textit{JWST}'s NIRSpec G235M/F170LP mode (1.66 - 3.07$\mu$m) would be capable of rejecting a flat-line spectrum at 5$\sigma$ confidence within 20 transits for a range of atmospheric compositions. While a flat-line rejection test is valuable, it does not guarantee the ability to detect the presence of any specific molecule with high statistical confidence.\cite{LustYeag19} concluded that transmission spectroscopy with NIRSpec PRISM is optimal for detecting the presence of high mean molecular weight atmospheres within $\sim$12 transits. Additionally, they find that CO$_{2}$, if present, will be easily detectable regardless of the atmospheric composition and MIRI-LRS may be capable of detecting the 9.6$\mu$m O$_3$ on planet 1e (with an SNR=3) with greater than 100 transits. \cite{Wunderlich19} produced self-consistent forward models for the atmosphere of an Earth-like planet around early-to-late M-dwarfs. Their ``Earth around TRAPPIST-1" model notably predicts a greatly increased temperature in the mid-to-upper atmosphere, inflating the scale height and increasing the detectability of molecular species for transmission spectroscopy with \textit{JWST}. Furthermore, the predicted chemical evolution indicated a significant enhancement of CH$_{4}$ and H$_{2}$O compared to modern Earth and subsequently, that their corresponding spectral features, along with CO$_{2}$, would be detectable within $\sim$10 transits when using the appropriate NIRSpec high-resolution filters.

\cite{Barstow16} performed a retrieval analysis on synthetic spectra of the inner-most TRAPPIST-1 companions (b, c, d) each with an Earth-like composition atmosphere. Utilizing a combination of NIRSpec PRISM and MIRI-LRS, they conclude that O$_{3}$, at Earth-like abundances, would be detectable on planets 1c and 1d with 30 transits on each instrument. Also utilizing an atmospheric retrieval analysis, \cite{KT18} analyzed the detectability of the CO$_{2}$+CH$_{4}$ disequilibrium biosignature, suspected to have been present in Earth's early atmosphere when the abundances of each were significantly enhanced compared to modern-day values. They concluded that \textit{JWST} would be capable of obtaining order-of-magnitude constraints on the mixing ratios of both CH$_{4}$ and CO$_{2}$ in an Archean Earth-like atmosphere, in $\sim$10 transits. Alternatively, the abundance of O$_{3}$ in a modern Earth-like atmosphere could not be constrained by NIRSpec Prism and would be completely unbounded by MIRI-LRS. 

While it may be possible for \textit{JWST} to detect and constrain the dominant molecular species in certain terrestrial atmospheres with advantageous compositions, existing literature does not substantiate its capabilities to detect or constrain the abundances of the five primary bio-indicators in a modern Earth-like atmosphere on an M-dwarf companion. 

The primary aim of this work is to develop a baseline understanding of how key observational parameters (spectral resolution, wavelength coverage, and signal-to-noise) of near-to-mid-infrared transmission spectra influence the ability to detect and constrain biologically relevant molecular species in the atmospheres of transiting Earth-like planets orbiting M-dwarfs.  We leverage powerful Bayesian retrieval tools to obtain both parameter constraints and Bayesian molecular detection significances as a function of the instrumental trades. In $\S$\ref{Modeling Techniques} we describe our model/retrieval and instrumental-trade space setup, $\S$\ref{Retrieval Results} summarizes the key results, followed by a discussion and key points in $\S$\ref{Discussion} and $\S$\ref{Conclusion}.  
 
\section{Simulation Set Up} \label{Modeling Techniques}
Here, we outline the development of our forward model used to compute the synthetic transmission spectra, elaborate on the instrumental trade space, discuss the details of our instrument noise model, and review our retrieval approach.

\subsection{Transmission Forward Model}
We simulate the transmission spectrum of a terrestrial planet, with mass and radius equivalent to TRAPPIST-1e, residing within the ``habitable zone" of TRAPPIST-1 with an approximate modern Earth-like atmosphere (observational system setup/signal-to-noise discussed below). We acknowledge, up front, that this may not be a physically plausible scenario due to the vastly different incident spectral energy distribution. However, given the overwhelming number of unknowns involved in self-consistent planetary atmosphere modeling (e.g., star-planet interactions, surface fluxes, formation conditions, bulk elemental composition, 3D atmospheric dynamical effects, cloud micro-physics, dynamical evolution/history) we simply choose to treat our atmosphere as ``Earth-like" (as has also been done in previous works, \cite{Morley17, KT18}).
 
  \begin{figure}
    \centering
    \includegraphics[width=\columnwidth, height=6.25cm]{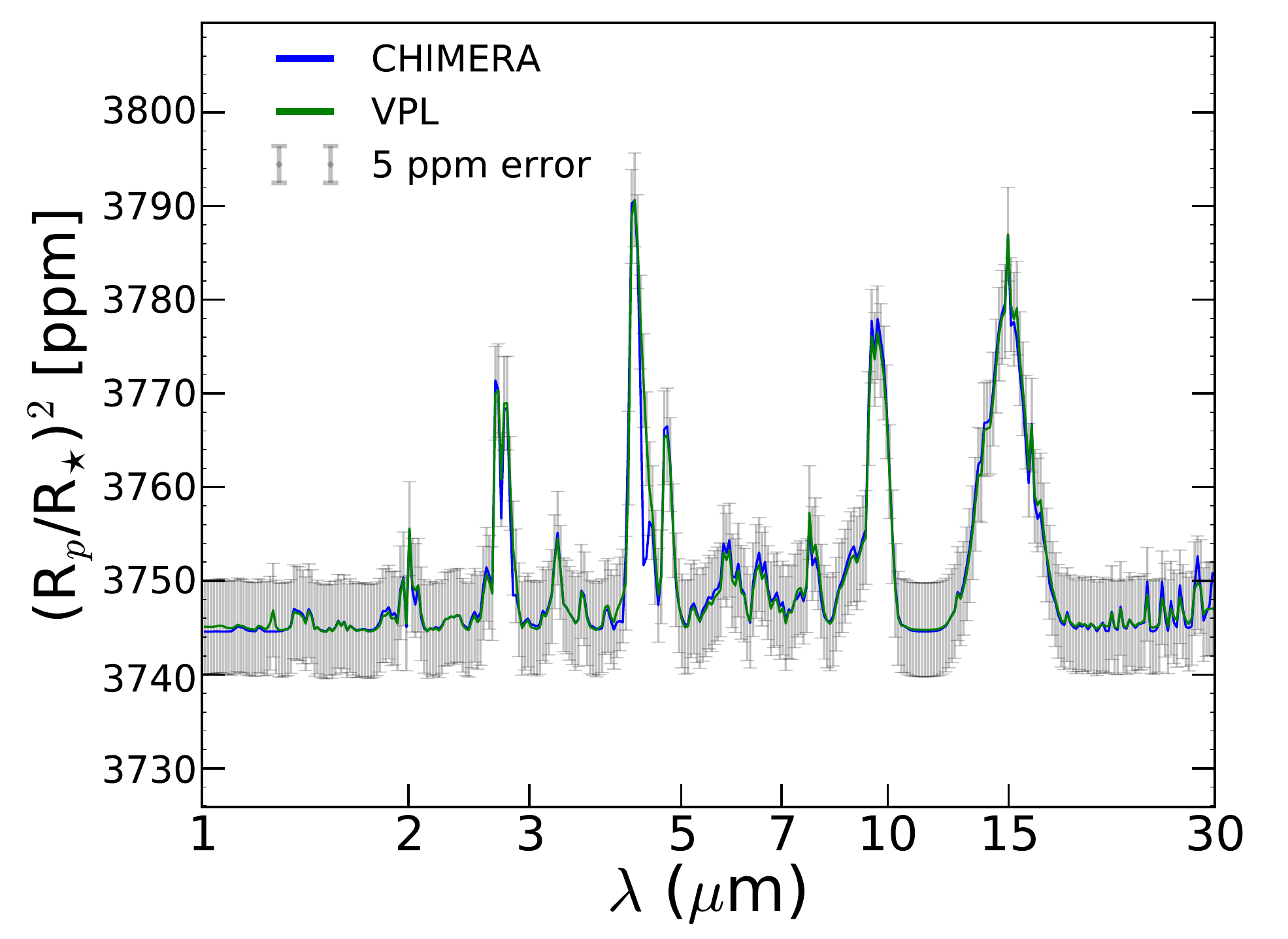}
    \caption{Validation of our transmission forward model (at R=100) against the Virtual Planetary Laboratory (VPL) Earth model$^3$ utilizing the same temperature-pressure and gas mixing ratio profiles. The spectra were scaled to a 1R$_{\Earth}$ around a 0.15R$_{\Sun}$ star. For reference, overlaid are 5 ppm error bars. }
    \label{validate}
\end{figure}
 
To produce model transmission spectra, we leverage a variant\footnote{\url{https://www.dropbox.com/sh/nehnyq6mlkfb13w/AABuDn-uFqTvAptWVSbGoA04a?dl=0}} of the {\tt CHIMERA}\footnote{\url{https://github.com/mrline/CHIMERA}} transmission spectrum routine (Batalha \& Line 2017; Line et al. 2013; Greene et al. 2016; Line \& Parmentier 2016) \citep{Batalha17,Line13,Greene16,Line16}.  Specifically, we re-parameterize the code (Table \ref{priors}) to make it more amenable for temperate worlds by including as free parameters the constant-with-altitude mixing ratios of H$_{2}$O, CO$_{2}$, CO, CH$_{4}$, O$_{3}$, N$_{2}$O and an ``unknown" background gas with a free-parameter molecular weight (taken to be earth's N$_2$+O$_2$ value), an isothermal ``scale-height" temperature, planetary radius at the surface (or scaling there-of), and an opaque gray cloud-top-pressure (nominal truth values given in Table \ref{priors}).  We mimic refraction with this cloud top pressure (here, 0.56 bars) using the prescription from \cite{Robinson17}, but otherwise assume a cloud free atmosphere (the refractive layer is in affect, mimicking a cloud, as well as its inclusion as a free parameter).  {\tt CHIMERA} uses correlated-K opacities (computed at the given constant resolving powers below), here derived from a grid of pre-computed line-by-line ($\le$0.01 cm$^{-1}$, 70 - 410 K, 10$^{-7}$ - 30 bar) cross-sections generated with the HITRAN HAPI Routine \citep{Kochonov16} and the HITRAN2016 line database \citep{Gordon17}. We do not include collision induced opacities, though they may spectrally present themselves throughout the mid-infrared \citep{Schwieterman15}.
 
 \begin{table}[ht]
\centering
\begin{tabular}{lcc}
\hline
\hline
               & Model Value      & Uniform Prior \\
\hline
$\log$(H$_{2}$O) & -5.5             & [-12, 0]      \\
$\log$(CO$_{2}$) & -3.45            & [-12, 0]      \\
$\log$(CH$_{4}$) & -6.3             & [-12, 0]      \\
$\log$(O$_{3}$)  & -6.5             & [-12, 0]      \\
$\log$(N$_{2}$O) & -6.3             & [-12, 0]      \\
$\log$(CO)     & -7.0             & [-12, 0]      \\
T$_{iso}$      & 280K             & [100, 800]    \\
xR$_{P}$       & 0.918R$_{\Earth}$ & [0.5, 1.5]    \\
$\log$(CTP)    & -0.25            & [-6, 2]       \\
Bkg MMW        & 28.6         & [2.0, 44.0] \\
\hline
\end{tabular}
\caption{The 10 free parameters, the model values, and their associated uniform priors used in our {\tt CHIMERA} radiative transfer and retrieval code.}
\label{priors}
\end{table}
 
We benchmarked our transmission forward model against those available from the Virtual Planetary Laboratory\footnote{\url{http://depts.washington.edu/naivpl/content/vpl-spectral-explorer})} (Figure \ref{validate}).

\newpage

\subsection{Parameter Estimation \& Model Selection Approach} \label{Retrieval Technique}
We perform Bayesian parameter estimation and model selection, on the simulated datasets described below, using the {\tt PyMultiNest} routine \citep{Buchner14} following the methods described in \cite{Benneke13}. We initialized our retrievals with 3000 live points. An advantage of nested sampling algorithms is ease of evidence computation, which can be used to assess model complexity. We use Bayesian nested model comparison (by removing one gas at a time) to determine the detection significance of each molecular species \citep{Trotta08,Benneke13}, and is what we utilize as a metric for assessing instrument performance. An advantage of utilizing the detection significance via the Bayesian evidence, over more traditional methods (e.g., “line” or “band” height above a continuum relative to the noise) is that it fully utilizes all of the spectral information included in all of the relevant bands as well as encompasses the model degeneracies.  

\subsection{Simulated Spectrograph \& Radiometric Noise Model} \label{Radiometric Noise Model}
We use the {\it JWST} radiometric noise model described in \cite{Greene16} to produce spectrophotometric uncertainties.
The collecting area of 25m$^2$ was retained to reflect the identical collecting area of \textit{JWST} in order to provide a baseline with which to easily scale the results of this study. Modifications were made to parameters of the code to reflect the expected performance of a next-generation mid-infrared telescope. These parameters included a 2.85 - 30.0 micron wavelength range (HgCdTe detectors below 10.5 microns and SiAs detectors above 10.5 microns), a zodiacal background estimated by \cite{Glass15a,Glass15b}, and an intrinsic resolving power of 300. These modifications reflect the capabilities of new detector technologies discussed by Matsuo et al. (2018).

Our input stellar spectrum derives from the PHOENIX stellar models using the {\tt PySynPhot} software package, \cite{Lim15}. We adopted the values for TRAPPIST-1 (T=2550K, M/H=0.40, $\log(g)$=4.0) and stellar radius of 81373.5km from \cite{Gillon16}. Additionally, we have scaled the stellar spectrum to a K-band magnitude of 8.0 rather than mag$_{K}=10.3$, TRAPPIST-1's native magnitude. While this may be representative of an optimistic scenario, mag$_{K}=8.0$ is the mean magnitude of the brightest 10 M-dwarf stars in the Barclay catalog of the predicted \textit{TESS} yield \citep{Barcklay18}.

\begin{table}[ht]
\centering
\begin{tabular}{ccc}
\hline
\hline
Short Wavelength      & Long Wavelength      & Spectral Resolution \\
Boundary             & Boundary          &       \\
\hline
\multirow{8}{*}{1$\mu$m} & \multirow{3}{*}{5$\mu$m}  & 300 \\ \cline{3-3} 
                         &                           & 100 \\ \cline{3-3} 
                         &                           & 50  \\ \cline{2-3} 
                         & \multirow{2}{*}{11$\mu$m} & 100 \\ \cline{3-3} 
                         &                           & 50  \\ \cline{2-3} 
                         & \multirow{3}{*}{30$\mu$m} & 100 \\ \cline{3-3} 
                         &                           & 50  \\ \cline{3-3} 
                         &                           & 30  \\ \hline
\multirow{8}{*}{3$\mu$m} & \multirow{3}{*}{5$\mu$m}  & 300 \\ \cline{3-3} 
                         &                           & 100 \\ \cline{3-3} 
                         &                           & 50  \\ \cline{2-3} 
                         & \multirow{3}{*}{11$\mu$m} & 300 \\ \cline{3-3}
                         &                           & 100 \\ \cline{3-3}
                         &                           & 50  \\ \cline{2-3} 
                         & \multirow{3}{*}{30$\mu$m} & 100 \\ \cline{3-3} 
                         &                           & 50  \\ \cline{3-3} 
                         &                           & 30  \\ \hline
\multirow{6}{*}{5$\mu$m} & \multirow{3}{*}{11$\mu$m} & 100 \\ \cline{3-3} 
                         &                           & 50  \\ \cline{3-3} 
                         &                           & 30  \\ \cline{2-3} 
                         & \multirow{3}{*}{30$\mu$m} & 100 \\ \cline{3-3} 
                         &                           & 50  \\ \cline{3-3} 
                         &                           & 30 \\
\hline
\end{tabular}
\caption{A breakdown of the 23 test cases evaluated in this study as a combination of wavelength coverage and spectral resolution.}
\label{test cases}
\end{table}

\begin{figure*}[h]
    \centering
    \includegraphics[width=\textwidth]{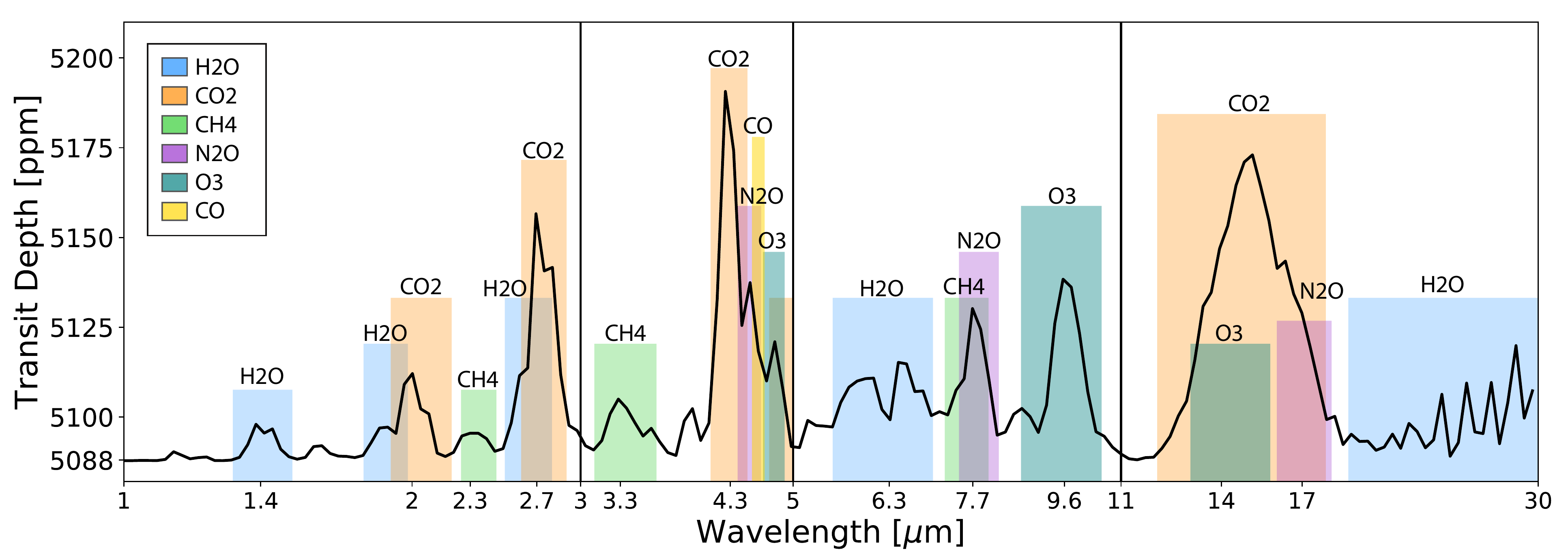}
    \caption{Wavelength region explored in this work (here, shown at R=50). The colored bands indicate the presence, and approximate width, of a spectral feature for a given molecule. We explore constraints obtainable over several wavelength bands within this region (in Table \ref{test cases}).}
    \label{stacked spectra}
\end{figure*}

\begin{figure*}[h]
    \centering
    \includegraphics[width=\textwidth]{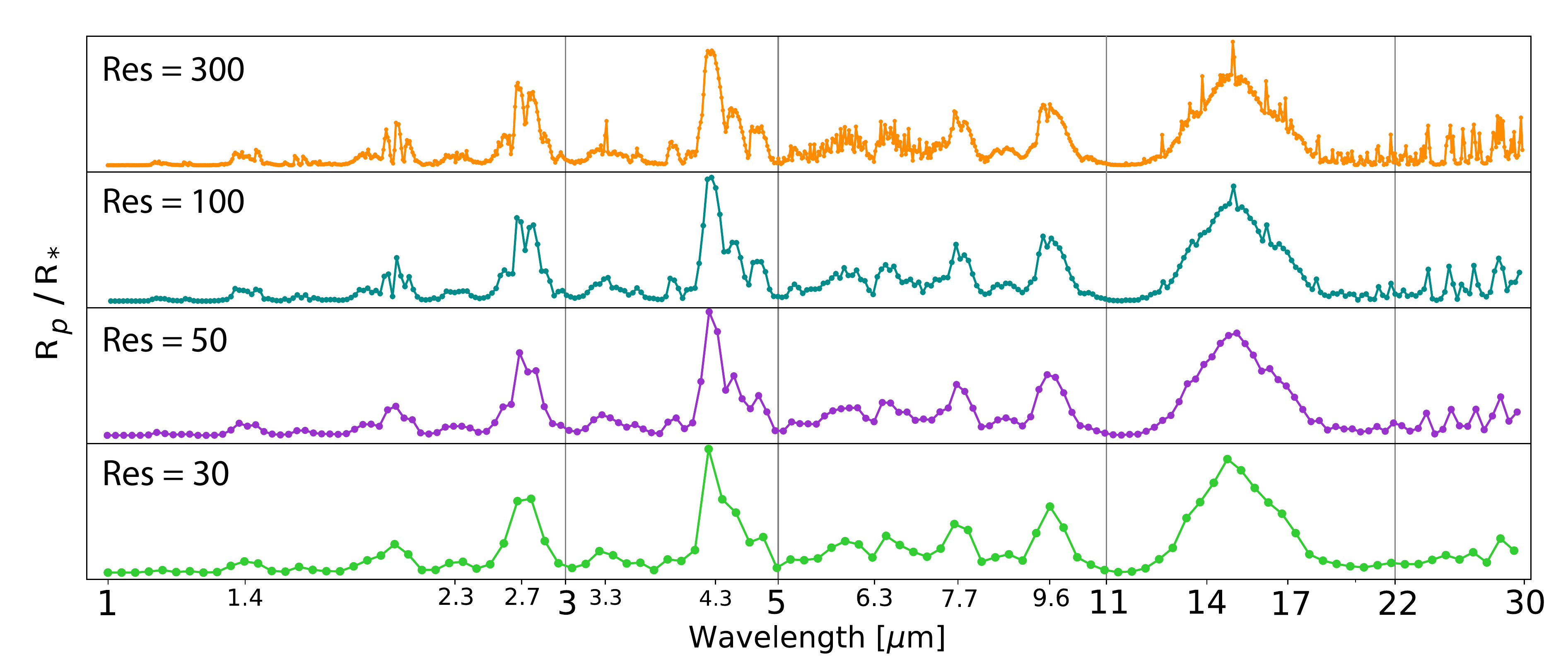}
    \caption{Influence of resolution on spectral features. Lower resolution results in the blending of features, causing a loss of information. The vertical lines at 3, 5, and 11$\mu$m denote the division of the wavelength regimes explored in this study.}
    \label{resolution compare}
\end{figure*}

\subsection{Observational Parameter Space}
We produce synthetic transmission spectra of our Earth-like atmosphere model over a grid of spectral resolution and bandpass choices. The nominal wavelength coverage was chosen to cover the near and mid-infrared regimes spanning 1-30$\mu$m, shown in Figure \ref{stacked spectra}. We divide this wavelength range into 8 separate regimes to explore the influence of additional bands of a given molecule and their overlap with features of other molecules. Additionally, we test low and moderate spectral resolutions ranging from R=30 to R=300. The spectral resolution ($\lambda / \delta\lambda$) dictates the degree to which individual molecular bands can be resolved and overlap degeneracies broken. Figure \ref{resolution compare} compares the spectra at the resolutions explored in this work. We simulate different resolutions within each spectral band in order to explore the resolution trade along with each wavelength regime. This produces a grid of 23 test cases (shown in Table \ref{test cases}) to evaluate through our retrieval algorithm described in \ref{Retrieval Technique}.

In addition to resolution and wavelength coverage, we also explored the signal-to-noise trade, parameterized here as the number of transits (1-100), which can be considered a proxy for mirror diameter or source magnitude, where the noise on a single transit is defined by the ``nominal" noise model setup described in \ref{Radiometric Noise Model}. Utilizing this trade, we explored the minimum number of transits necessary to achieve a threshold 3.6$\sigma$ confidence level detection of each molecule. We provide equations \ref{Transit conversion mag} \& \ref{Transit conversion area} as a means of converting our retrieval results to stars of differing magnitudes or for different sized collecting mirrors. 

\begin{equation} \label{Transit conversion mag}
    M_{eff} = M_{nom} - 2.5\log_{10}(\frac{T_{eff}}{T_{nom}})
\end{equation}
Where $T_{eff}$/$T_{nom}$ is the ratio determining how many more or less transits are equivalent to a change from the native magnitude ($M_{nom}=8.0$) to a new magnitude ($M_{eff}$).
\begin{equation} \label{Transit conversion area}
    A_{eff} = A_{nom}\sqrt{\frac{T_{eff}}{T_{nom}}}
\end{equation}
Where $T_{eff}$/$T_{nom}$ is the ratio of transits equivalent to a change from the native collecting area ($A_{nom}=25$m$^2$) to a new mirror size ($A_{eff}$).

\section{Retrieval Results} \label{Retrieval Results}
For the majority of our test cases, the inclusion of a broader bandpass decreases both the necessary observation time to claim a statistically significant detection and reduces the uncertainty on the abundance constraints. Therefore, with few molecule-specific exceptions, we find that altering the choice of wavelength coverage has a more prominent benefit for both our detection significance values and abundance constraints than the choice of spectral resolution (Tables  \ref{detection significance summary} \& \ref{abundance constraint summary}). 

Our detection significance analysis directly compares each bandpass and resolution combination by determining the number of transits required to achieve a 3.6$\sigma$ detection for a specific molecule. Table \ref{detection significance summary} displays these required number of transits for each molecule over each combination of resolution and wavelength coverage. 

\begin{figure}[h]
    \centering
    \includegraphics[width=\columnwidth]{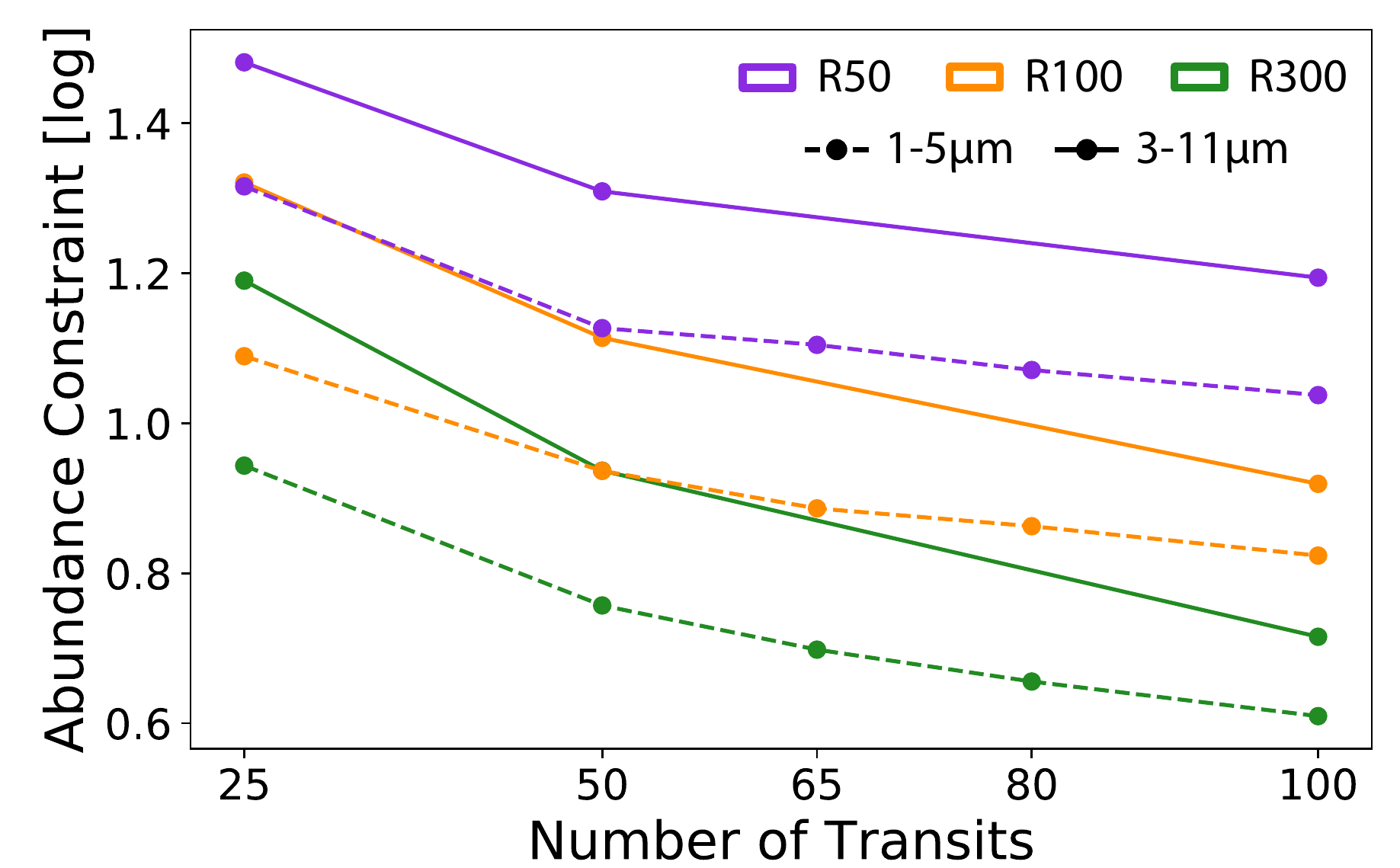}
    \caption{Compares the improvements of the abundance constraints for H$_2$O, with increasing numbers of transits at three resolutions for the two key bandpass choices.}
    \label{constraint trend H2O}
\end{figure}

\begin{figure}[h]
    \centering
    \includegraphics[width=\columnwidth]{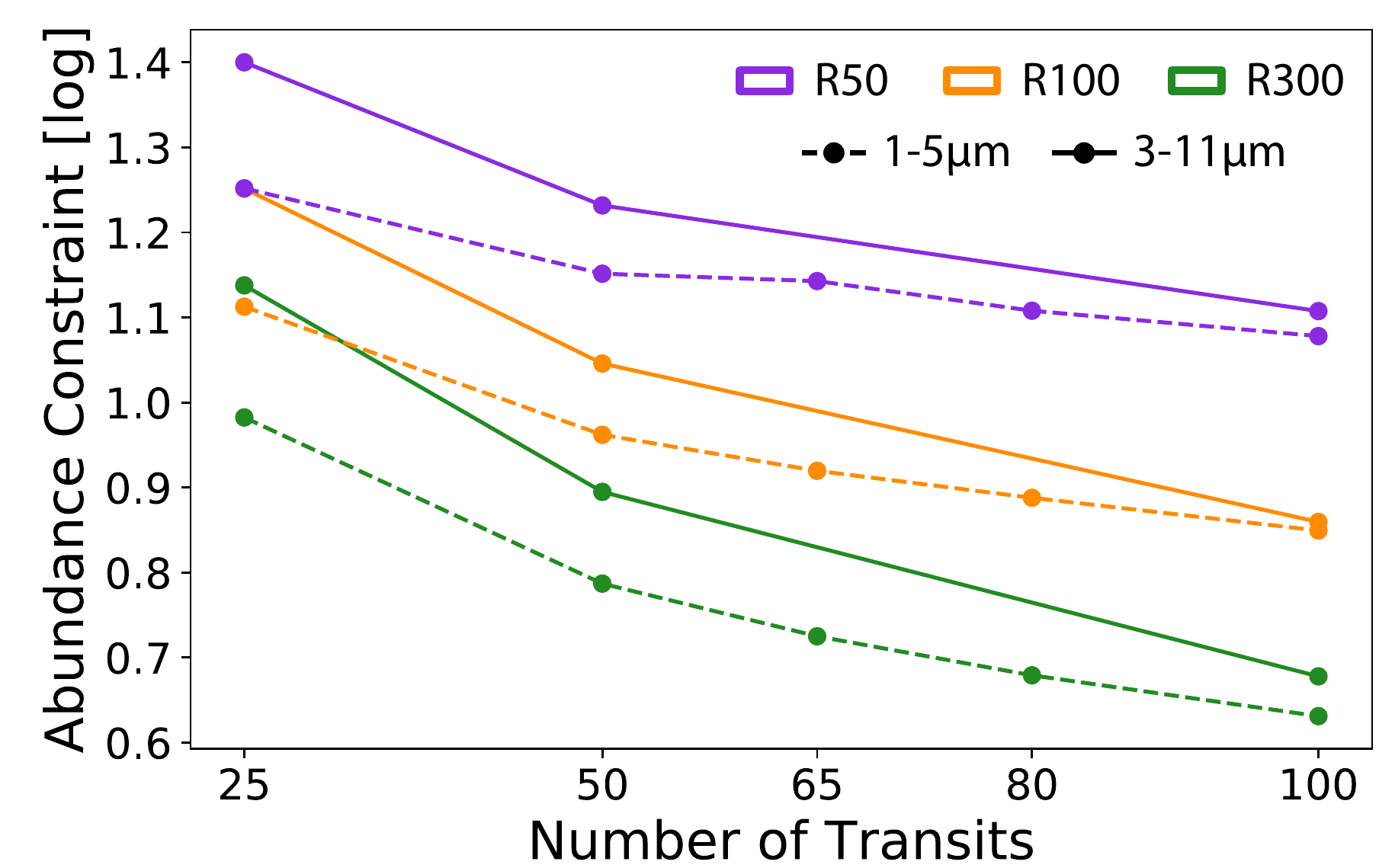}
    \caption{Compares the improvements of the abundance constraints for CO$_2$, with increasing numbers of transits at three resolutions for the two key bandpass choices.}
    \label{constraint trend CO2}
\end{figure}

\begin{figure}[h]
    \centering
    \includegraphics[width=\columnwidth]{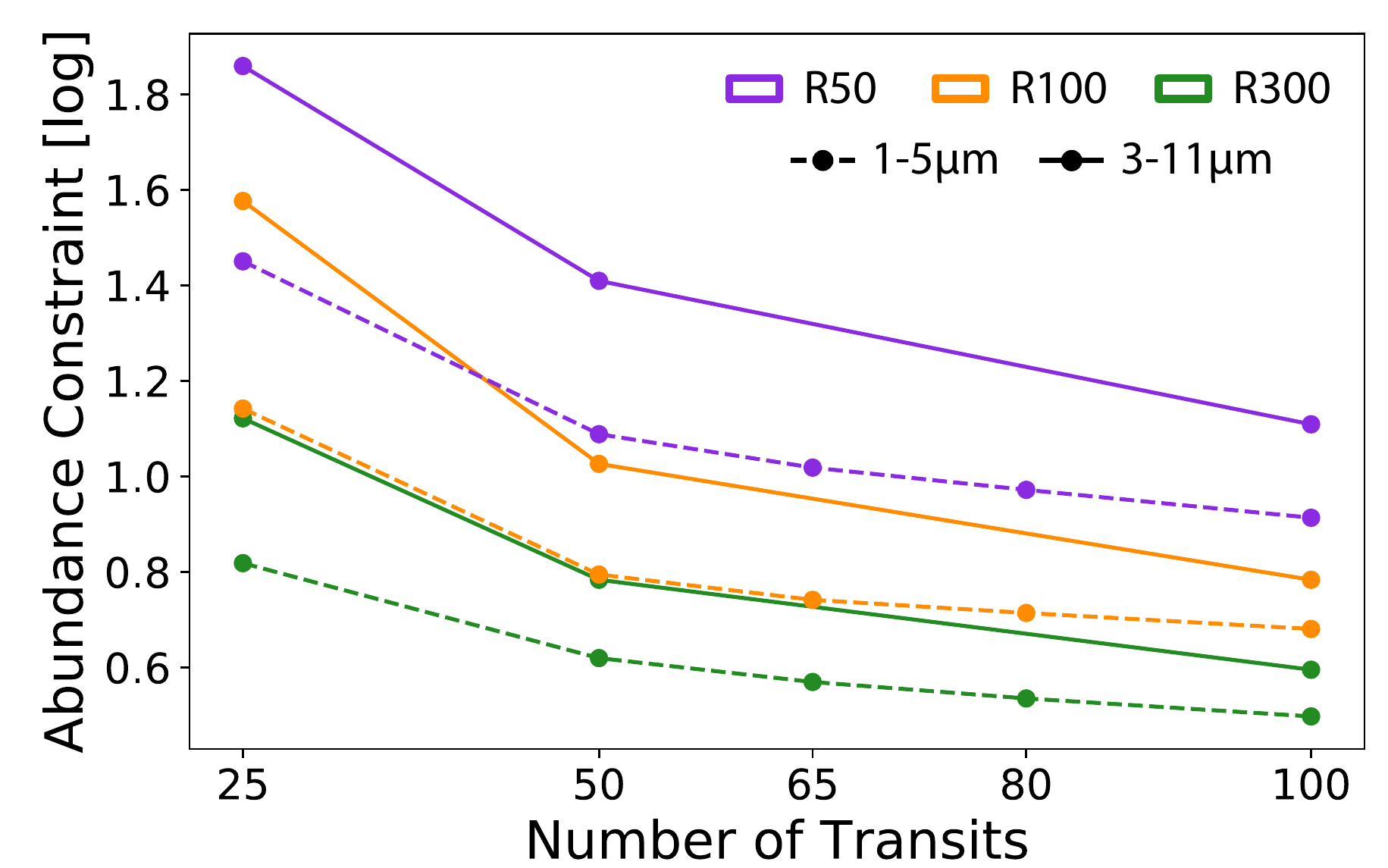}
    \caption{Compares the improvements of the abundance constraints for CH$_4$, with increasing numbers of transits at three resolutions for the two key bandpass choices.}
    \label{constraint trend CH4}
\end{figure}

\begin{figure}[h]
    \centering
    \includegraphics[scale=0.425]{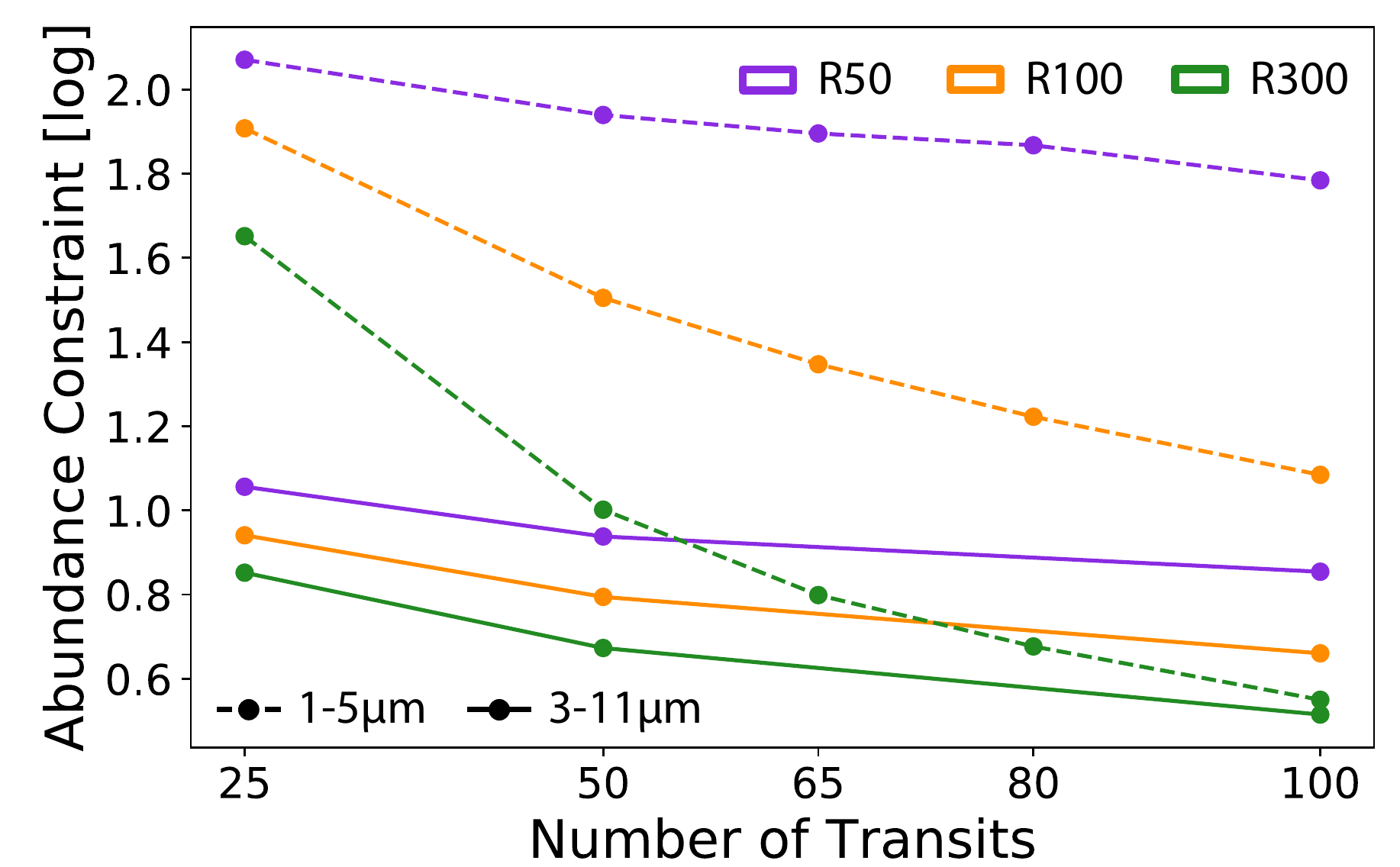}
    \caption{Compares the improvements of the abundance constraints for O$_3$, with increasing numbers of transits at three resolutions for the two key bandpass choices.}
    \label{constraint trend O3}
\end{figure}

\begin{figure}[ht]
    \centering
    \includegraphics[width=\columnwidth]{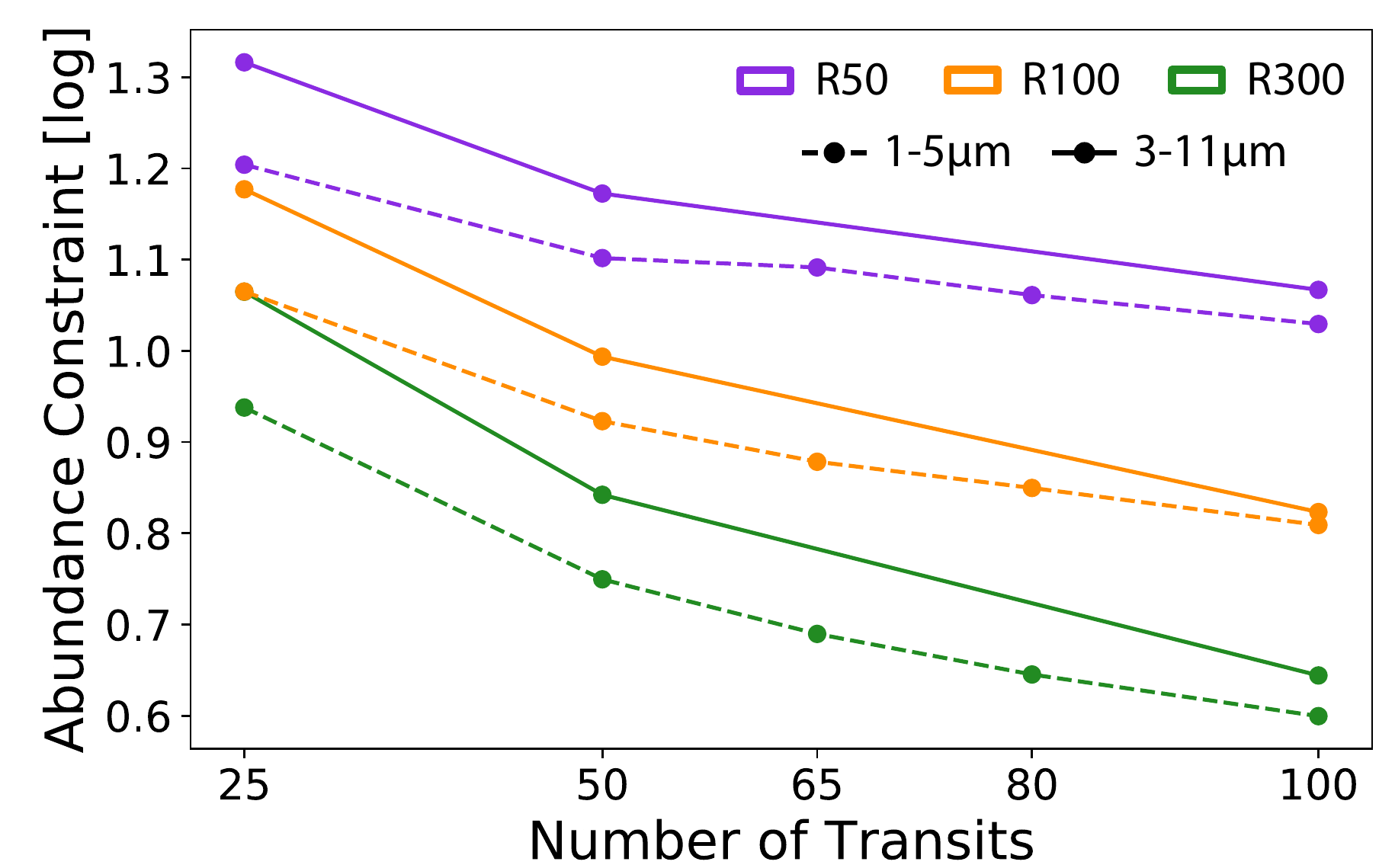}
    \caption{Compares the improvements of the abundance constraints for N$_2$O, with increasing numbers of transits at three resolutions for the two key bandpass choices.}
    \label{constraint trend N2O}
\end{figure}

Table \ref{abundance constraint summary} enumerates the 1$\sigma$ error widths for each molecule within each of our 23 test cases at 50 transits. Figures \ref{constraint trend H2O},\ref{constraint trend CO2},\ref{constraint trend CH4},\ref{constraint trend O3}, and \ref{constraint trend N2O} within each of the following sections reveal the influence that an increasing number of transits has on the abundance constraints, for each molecule individually at R=100. The following subsections provide an in-depth analysis of each molecule and the most relevant transitions that influence our abundance constraints and detection significance results.

\subsection{Water (H$_{2}$O)}

\begin{figure*}[h]
    \centering
    \includegraphics[scale=0.6]{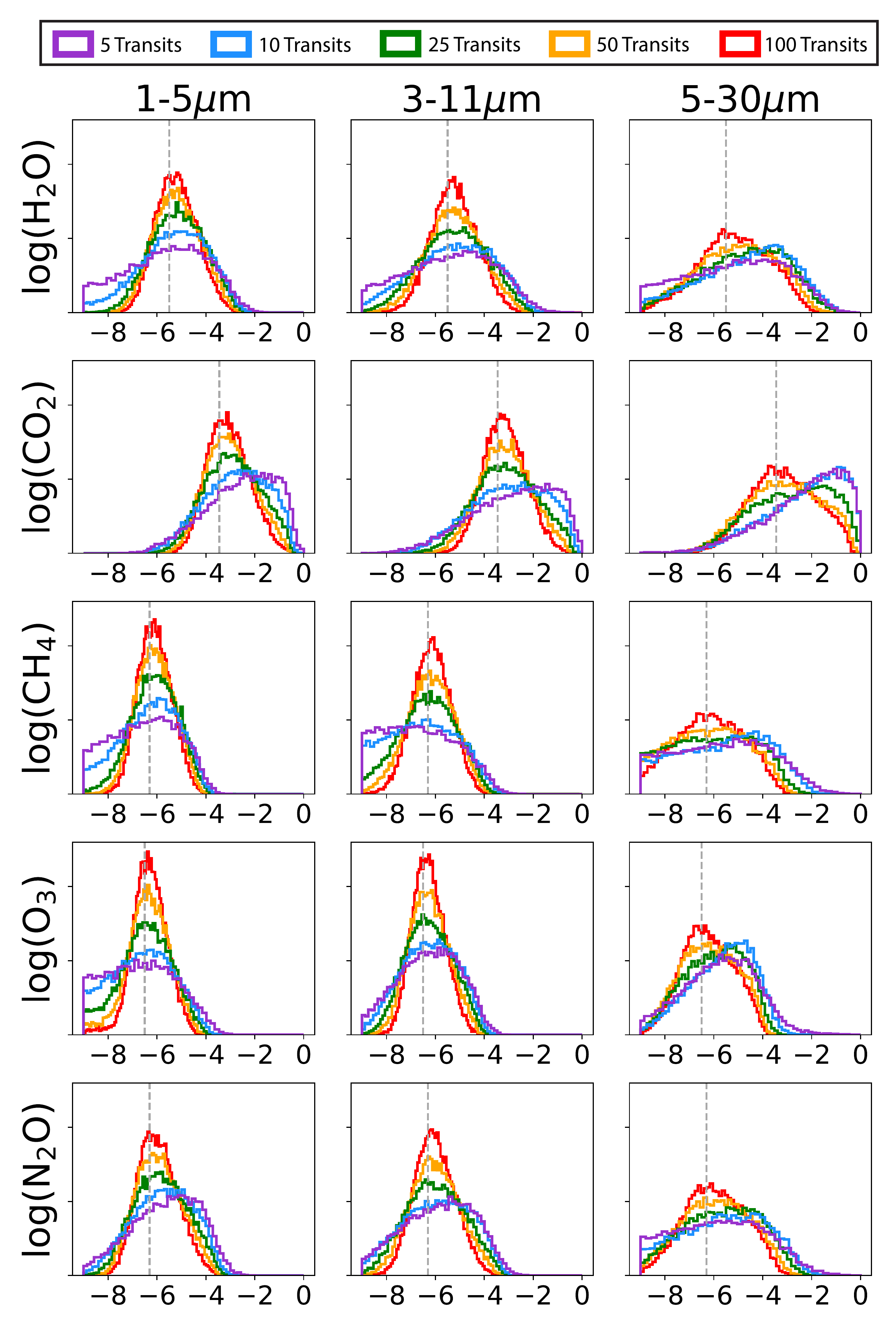}
    \caption{Abundance constraints for each molecule within three key bandpasses (1-5$\mu$m, 3-11$\mu$m, and 5-30$\mu$m at R=100) as a function of the number of observed transits. The heights of each histogram window has been fixed to a constant value to allow for a direct comparison of the shape of the marginalized posterior probability distributions for each gas. The ``hard edge" of some distributions, near low abundances, indicates those that extend beyond the prior range.}
    \label{block hist}
\end{figure*}

Water has spectral features spanning nearly the entire wavelength range in this study, with the most prominent features centered at 2.65$\mu$m and 6.3$\mu$m. Additionally, there are two smaller features extending into the near-infrared at 1.4$\mu$m and 1.85$\mu$m as well as a series of features upwards of 20$\mu$m. The most relevant comparison is between the near-infrared (which we define as the 1-5$\mu$m bandpass) and the mid-infrared (5-11$\mu$m). In isolation from other water features in the 5-11$\mu$m region, the broad 6.3$\mu$m feature produces comparatively underwhelming results (for both detection significance and abundance constraints) likely due to the limited contribution from the few other prominent features in this bandpass \citep{Benneke12}. In contrast, the 1-5$\mu$m region encompasses the prominent 2.65$\mu$m feature as well as the two smaller features. The precision on the abundance constraints improves by approximately 60\% both when increasing the resolution from R=50 to R=100 and again from R=100 to R=300. However, it is worth noting that these near-infrared features are very narrow and are thus significantly more sensitive to changes in the choice of spectral resolution. At a resolution of R=100, a single H$_{2}$O feature in the near-infrared consists of between 10-20 resolution elements compared to 44 for the 6.3$\mu$m feature. 

While the near-infrared (1-5$\mu$m bandpass) produces the tightest abundance constraints, it does not provide the greatest detection significance values given the same spectral resolution and number of transits (refer to Table 3 for comparison of detection significance results). By comparison, the 3-11$\mu$m range (which does not include additional H$_{2}$O features) requires 33.9\% less observation time at R=100 and 50.0\% less at R=50 than the 1-5$\mu$m bandpass. Alternatively, by looking further into the mid-infrared, the 5-30$\mu$m bandpass offers a 3.4\% decrease at R=100 and a 24.5\% decrease at R=50 compared to the 1-5$\mu$m bandpass.

\subsection{Carbon Dioxide (CO$_{2}$)}

Due to its prominent spectral features across most of the near and mid-infrared, CO$_{2}$ is the easiest molecule to detect within each of our bandpass choices, with the exception of 5-11$\mu$m. CO$_{2}$ possess four prominent spectral features centered at 2.0, 2.7, 4.3, and 15um. The 5-11$\mu$m region is the only bandpass choice which does not include a spectral feature for CO$_{2}$. Any other choice of wavelength coverage and spectral resolution evaluated in our study proved to be sufficient to detect CO$_{2}$ in less than 7 transits. The inclusion of the 4.3$\mu$m feature ensures a 3.6$\sigma$ detection within the observation time of approximately a single transit. 

Despite the strong detection of CO$_{2}$, the abundance constraints are ironically, not as precise as would be expected. This is largely due to the fact that deriving narrow abundance constraints relies on a change in the strengths of the spectral features.
\cite{Line16} derive the function for spectral modulation with respect to wavelength for multiple absorbing molecular species, reproduced here for convenience (Equation \ref{spectral modulation}).

\begin{equation} \label{spectral modulation}
    \frac{d\alpha_{\lambda}}{d\lambda}=\frac{2R_{p}}{R_{\star}^{2}}H\frac{1}{1+\frac{\xi_{2}\sigma_{\lambda,2}}{\xi_{1}\sigma_{\lambda,1}}}\left(\frac{d\ln(\sigma_{\lambda,1})}{d\lambda}+\frac{\xi_{2}\sigma_{\lambda,2}}{\xi_{1}\sigma_{\lambda,1}}\frac{d\ln(\sigma_{\lambda,2})}{d\lambda}\right)
\end{equation}

Where d$\alpha_{\lambda}$/d$\lambda$ is the wavelength dependent slope of the transmission spectra. $H$ is the scale height given by $\frac{k_{b}T}{\mu g}$ and $\sigma_{\lambda,i}$ and $\xi_{i}$ are the absorption cross section and abundance, respectively, of a given molecular species. 

In the case of CO$_{2}$, the $\xi_{1}\sigma_{\lambda,1}$ term is so much greater than the other absorbers that the spectral modulation due to CO$_{2}$ becomes insensitive to even large changes in the abundance. The retrieval thus determines that a wide range of abundance values can reproduce the shape of the observed features within the error bars. Therefore, the only means to narrow the constraint is to increase the number of observed transits. 

At 50 transits, the 1-5$\mu$m bandpass produces approximately a single order-of-magnitude (1.01dex) constraint at R=100 or 0.81dex at R=300. The abundance constraints produced by the 1-5$\mu$m bandpass cases at both R=50 and R=100 are comparable to those of both the 3-11$\mu$m and 3-30$\mu$m bandpasses.

\subsection{Methane (CH$_{4}$)}

\begin{figure}[h]
    \centering
    \includegraphics[width=\columnwidth]{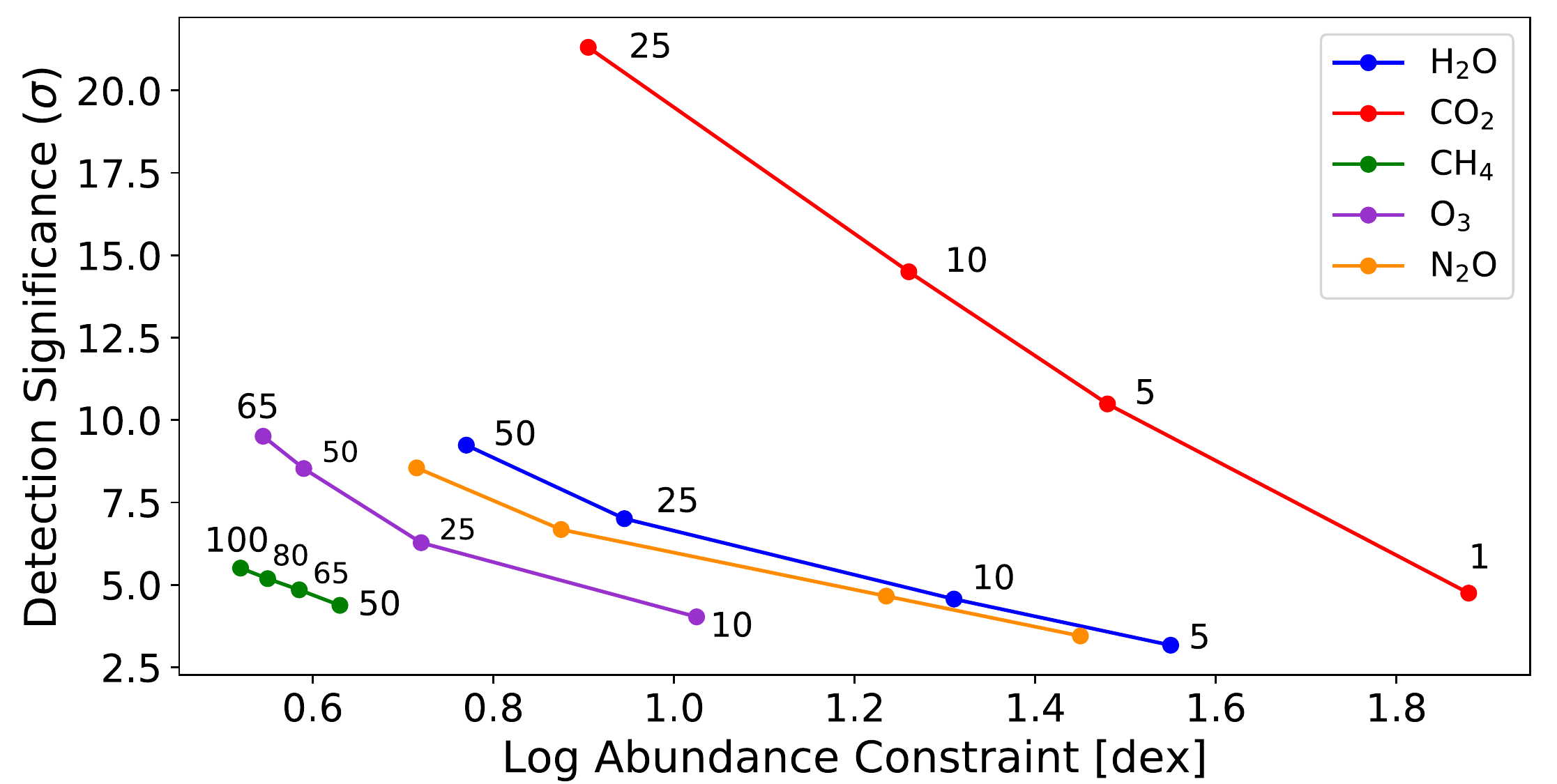}
    \caption{The relationship between detection significance and abundance constraint. The numbers on each data point represent the number of transits observed to achieve those respective results. The number of transits for the N$_2$O (orange) are identical to those of H$_2$O (blue), conveniently adjacent here. These results are representative of an instrument with 1-30$\mu$m coverage,  R=100 for a stellar m$_K$=8 and mirror aperture of 25m$^2$). Unsurprisingly, an increase in detection significance is correlated with higher precision. However, the mapping between detection significance and constraint is molecule specific.}
    \label{DSvAC}
\end{figure}

Due to its low abundance in the modern Earth atmosphere, CH$_{4}$ is difficult to detect and ultimately determines the lower limit on the observing time required to detect all five bio-indicator molecules. Despite the larger number of transits required to detect CH$_{4}$, we can place the best constraints on its abundance. Indeed, it is notable that our results indicate that for the 3-11$\mu$m at R=100 case, a 3.6$\sigma$ detection is only possible after 61.7 observed transits while at only 50 transits we can place an order-of-magnitude constraint on its abundance. This appears initially paradoxical, until we examine the underlying dependencies of these two different results. Effectively, the detectability (via the Bayes factor) depends on the amplitude of the spectral features with respect to the continuum or adjacent features of other molecules (e.g., wavelength dependent derivative of transit depth shown in equation \ref{spectral modulation}). The mixing ratio constraints on the other hand are dictated by the derivative of the transit depth with respect to a given species abundance (or rather, the inverse, e.g., \cite{Line12}), with each species having varying sensitivity. Therefore, one could observe small features resulting in a low detectability (as in the case of CH$_4$) while having a large abundance derivative, resulting in a tight constraint. Figure \ref{DSvAC} illustrates the  molecule-specific relationship between the abundance constraints (log) and detection significance under the 1-30$\mu$m, R=100 scenario for varying numbers of transits. Figure \ref{DSvAC} illuminates the key differences between CH$_{4}$ and CO$_{2}$ in that while CO$_{2}$ has a high detection significance, it does not necessarily have a precise constraint. Alternatively, even low confidence detections of CH$_{4}$ can yield precise abundance constraints.  

At Earth-like abundances, CH$_{4}$ has only three discernible spectral features in the near and mid-infrared: 2.3, 3.3, and 7.6$\mu$m. The most prominent of these is centered at 7.6$\mu$m and directly overlaps with a larger N$_{2}$O feature centered at 7.7$\mu$m. Bandpass choices that exclude additional methane features will lack the ability to overcome this degeneracy. Fortunately, the 3.3$\mu$m ($\nu_{3}$) feature provides an unambiguous marker for the presence of CH$_{4}$. We find that the inclusion of this feature is crucial to detecting CH$_{4}$ as neither the 5-11$\mu$m nor the 5-30$\mu$m bandpasses are capable of detecting methane within 100 transits, despite the presence of the 7.6$\mu$m ($\nu_{4}$) feature. There is an additional feature at 2.3$\mu$m which, when combined with the 3.3$\mu$m band, greatly improves the detection significance. 

When observing only the 3-5$\mu$m bandpass, at R=300, CH$_{4}$ is detectable with 54.9 transits and provides an abundance constraint of 1.04dex at 50 transits. Maintaining this narrow wavelength coverage and reducing to R=100, increases the necessary observation time to beyond 100 transits. However, expanding our bandpass either to the near or mid-infrared regimes proves to provide significant benefits. The near-infrared bandpass (1-5$\mu$m) requires 38.5 transits and provides a 0.63dex (at 50 transits) abundance constraint or 54.4 transits and 0.81dex for R=300 and R=100, respectively. Alternatively, observing the 3-11$\mu$m bandpass requires 34.3 transits and provides a 0.79dex (at 50 transits) abundance constraint or 61.7 transits and 0.99dex for R=300 and R=100, respectively. 

Based on these results, we find that including the $\nu_{3}$ feature, within the chosen bandpass, is essential to detecting and constraining CH$_{4}$. In the scenario that spectral resolution is limited to R=100, the addition of the 2.3$\mu$m feature is incredibly beneficial, providing a 0.5dex improvement in the abundance constraint and requiring only a third of the observational time. However, while it is certainly ideal to include both features if possible, it is quite noteworthy that the benefit of a higher resolution (R=300) choice, in the absence of the 2.3$\mu$m feature, outweighs the benefit of including this feature at a lower resolution (R=100).

\begin{table}[h]
    \centering
    \begin{tabular}{|c|c|c|c|c|c|c|}
    \hline
Bandpass                    & Resolution & H$_{2}$O & CO$_{2}$ & CH$_{4}$ & O$_{3}$ & N$_{2}$O \\ \hline
\multirow{3}{*}{1-5$\mu$m}  & 300        & \textcolor{blue}{18.1}     & $<$ 1    & \textcolor{blue}{38.5}     & -       & \textcolor{blue}{7.7}      \\ \cline{2-7} 
                            & 100        & 23.3     & $<$ 1    & \textcolor{blue}{54.4}     & -       & 9.2      \\ \cline{2-7} 
                            & 50         & 39.2     & $<$ 1    & -        & -       & 14.3     \\ \hline
\multirow{2}{*}{1-11$\mu$m} & 100        & 6.5      & $<$ 1    & 44       & 8.5     & 5.7      \\ \cline{2-7} 
                            & 50         & 7.3      & $<$ 1    & 52.1     & 9.4     & 7.4      \\ \hline
\multirow{3}{*}{1-30$\mu$m} & 100        & 6.3      & $<$ 1    & 43.8     & 8.3     & 5.5      \\ \cline{2-7} 
                            & 50         & 7.0      & $<$ 1    & 50       & 9.1     & 7.2      \\ \cline{2-7} 
                            & 30         & 8.5      & 1.13     & -        & 10.9    & 9.0      \\ \hline
\multirow{3}{*}{3-5$\mu$m}  & 300        & -        & 1.03     & 54.9     & -       & 11.2     \\ \cline{2-7} 
                            & 100        & -        & 1.08     & -        & -       & 14.4     \\ \cline{2-7} 
                            & 50         & -        & 1.13     & -        & -       & 23.8     \\ \hline
\multirow{3}{*}{3-11$\mu$m} & 300        & \textcolor{blue}{11.9}     & $<$ 1    & \textcolor{blue}{34.3}     & \textcolor{blue}{8.9}     & \textcolor{blue}{5.6}      \\ \cline{2-7} 
                            & 100        & \textcolor{blue}{15.4}     & 1.07     & 61.7     & \textcolor{blue}{10.5}    & \textcolor{blue}{7.9}      \\ \cline{2-7} 
                            & 50         & 19.6     & 1.11     & -        & \textcolor{blue}{12.9}    & 10.6     \\ \hline
\multirow{3}{*}{3-30$\mu$m} & 100        & 14.4     & $<$ 1    & 60       & 10.2    & 7.7      \\ \cline{2-7} 
                            & 50         & 17.4     & $<$ 1    & -        & 11.8    & 9.6      \\ \cline{2-7} 
                            & 30         & 21.1     & 1.15     & -        & 14.7    & 12.2     \\ \hline
\multirow{3}{*}{5-11$\mu$m} & 100        & 33.1     & -        & -        & 14.9    & 94.0     \\ \cline{2-7} 
                            & 50         & 50       & -        & -        & 18.2    & -        \\ \cline{2-7} 
                            & 30         & 67.2     & -        & -        & 24.3    & -        \\ \hline
\multirow{3}{*}{5-30$\mu$m} & 100        & 22.5     & 6.75     & -        & 12.9    & 72.8     \\ \cline{2-7} 
                            & 50         & 29.6     & 6.88     & -        & 15.1    & -        \\ \cline{2-7} 
                            & 30         & 36.8     & 6.96     & -        & 18.2    & -        \\ \hline
    \end{tabular}
    \caption{Shown are the number of transits required to detect each molecule to 3.6$\sigma$ confidence for each of our test cases. The values in blue indicate the three lowest number of transits (for each molecule with the exception of CO$_2$) from the shortest wavelength ranges. A dash indicates that achieving a 3.6$\sigma$ was not possible within 100 transits and therefore, outside the scope of this study.}
    \label{detection significance summary}
\end{table}
\begin{table}[h]
    \centering
    \begin{tabular}{|c|c|c|c|c|c|c|}
    \hline
Bandpass                    & Resolution & H$_{2}$O & CO$_{2}$ & CH$_{4}$ & O$_{3}$ & N$_{2}$O \\ \hline
\multirow{3}{*}{1-5$\mu$m}  & 300        & \textcolor{blue}{0.77}     & \textcolor{blue}{0.81}     & \textcolor{blue}{0.63}     & \textcolor{blue}{0.65}       & \textcolor{blue}{0.77}     \\ \cline{2-7} 
                            & 100        & \textcolor{blue}{0.99}     & 1.01     & \textcolor{blue}{0.81}     & $\uparrow$ & \textcolor{blue}{0.97}     \\ \cline{2-7} 
                            & 50         & 1.19     & 1.24     & 1.03     & $\uparrow$ & 1.19     \\ \hline
\multirow{2}{*}{1-11$\mu$m} & 100        & 0.79     & 0.77     & 0.65     & 0.60       & 0.74     \\ \cline{2-7} 
                            & 50         & 0.99     & 0.93     & 0.81     & 0.75       & 0.92     \\ \hline
\multirow{3}{*}{1-30$\mu$m} & 100        & 0.77     & 0.74     & 0.63     & 0.59       & 0.72     \\ \cline{2-7} 
                            & 50         & 0.97     & 0.91     & 0.80     & 0.73       & 0.89     \\ \cline{2-7} 
                            & 30         & 1.15     & 1.08     & 0.99     & 0.87       & 1.07     \\ \hline
\multirow{3}{*}{3-5$\mu$m}  & 300        & $\uparrow$ & 1.31     & 1.04     & 0.95       & 1.16     \\ \cline{2-7} 
                            & 100        & $\uparrow$ & 1.44     & 1.32     & $\uparrow$ & 1.27     \\ \cline{2-7} 
                            & 50         & $\uparrow$ & 1.86     & $\uparrow$ & $\uparrow$   & 1.65     \\ \hline
\multirow{3}{*}{3-11$\mu$m} & 300        & \textcolor{blue}{0.95}     & \textcolor{blue}{0.91}     & \textcolor{blue}{0.79}     & \textcolor{blue}{0.68}       & \textcolor{blue}{0.85}     \\ \cline{2-7} 
                            & 100        & 1.15     & 1.07     & \textcolor{blue}{0.99}     & \textcolor{blue}{0.82}       & 1.02     \\ \cline{2-7} 
                            & 50         & 1.37     & 1.29     & 1.30     & \textcolor{blue}{0.99}       & 1.23     \\ \hline
\multirow{3}{*}{3-30$\mu$m} & 100        & 1.10     & 0.74     & 0.94     & 0.79       & 0.99     \\ \cline{2-7} 
                            & 50         & 1.33     & 0.91     & 1.21     & 0.96       & 1.21     \\ \cline{2-7} 
                            & 30         & 1.54     & 1.08     & 1.52     & 1.10       & 1.38     \\ \hline
\multirow{3}{*}{5-11$\mu$m} & 100        & 1.89     & -        & $\uparrow$ & 1.41       & 1.70     \\ \cline{2-7} 
                            & 50         & 2.05     & -        & -        & 1.58       & 2.01     \\ \cline{2-7} 
                            & 30         & 2.10     & -        & -        & 1.58       & 2.03     \\ \hline
\multirow{3}{*}{5-30$\mu$m} & 100        & 1.70     & 1.61     & $\uparrow$ & 1.23       & 1.50     \\ \cline{2-7} 
                            & 50         & 1.80     & 1.68     & $\uparrow$ & 1.29       & 1.69     \\ \cline{2-7} 
                            & 30         & 1.92     & 1.74     & $\uparrow$ & 1.33       & 1.78     \\ \hline
    \end{tabular}
    \caption{Shown are the logarithmic constraints on the vertical mixing ratios for each molecule, at 50 transits, for each of our test cases. The values reported are the half-width of the posterior distribution, therefore, 1.0 dex indicates the ability to constrain the abundance to within an order-of-magnitude higher or lower (e.g., $\pm$1 dex) than the retrieved value. The values in blue indicate the constraints better than one order-of-magnitude (for each molecule) from the shortest wavelength ranges. The $\uparrow$ marker denotes an upper limit and dash represents that the parameter was entirely unconstrained by the retrieval. The posteriors for all of the R=50 and R=100 test cases will be provided as supplementary material available at: \url{https://www.dropbox.com/sh/nehnyq6mlkfb13w/AABuDn-uFqTvAptWVSbGoA04a?dl=0}}
    \label{abundance constraint summary}
\end{table}

\subsection{Ozone (O$_{3}$)}

Ozone is different from the other molecular species we have addressed thus far, in that its most prominent features extend further into the mid-infrared. The three primary features for O$_{3}$ are at 4.75, 9.6, and 14.2$\mu$m with the 9.6$\mu$m feature being the strongest. Conveniently, due to the location of these spectral features and our bandpass choices, we can evaluate the benefit of each additional feature, individually, on the retrieval results. 

Isolating the 9.6$\mu$m feature, in the 5-11$\mu$m bandpass, requires a relatively low number of transits at only 14.9 and 18.2 transits for resolutions of R=100 and R=50, respectively. However, the abundance constraints for this bandpass exceed an order-of-magnitude at 1.41dex (R=100) and 1.58dex (R=50). If we choose to observe further into the mid-infrared (5-30$\mu$m), including the features at 9.6$\mu$m and 14.7$\mu$m, we notice only slight reductions in the required observation time and maintain larger than an order-of-magnitude constraint on both abundance values. The one noticeable benefit of including these longer wavelengths is that less than 20 transits are required to detect O$_{3}$ even at a resolution of R=30. Alternatively, extending the bandpass to 3-11$\mu$m to include the 4.75$\mu$m feature, results in a reduction of the necessary observation time to 10.5 transits for a resolution of R=100 (12.9 for R=50) and improves our abundance constraints by nearly a factor of four (0.82dex and 0.99dex) for both R=100 and R=50. It is worth noting that the 4.75$\mu$m feature has significant overlap with a minor CO$_{2}$ feature, thereby causing a degeneracy if observed without additional O$_{3}$ features. We conclude that the optimal wavelength coverage for detecting and constraining O$_{3}$ is the 3-11$\mu$m bandpass choice. Additionally, this bandpass would only require a resolution of R=50 to place an order-of-magnitude constraint on the volume mixing ratio for O$_{3}$.

\subsection{Nitrous Oxide (N$_{2}$O)}

Similar to O$_{3}$, the most prominent features for N$_{2}$O extend out into the longer wavelengths. N$_{2}$O's strongest spectral features are located at 4.5$\mu$m and 7.7$\mu$m, each with an adjacent smaller feature at 3.9$\mu$m and 8.6$\mu$m, respectively. Although there is a broad feature at 17$\mu$m it is overlapped by a much larger CO$_{2}$ feature. 

Due to the large degree of degeneracy with the 7.6$\mu$m CH$_{4}$ feature, isolating the 7.7$\mu$m and 8.6$\mu$m features in the 5-11$\mu$m bandpass proves to be inadequate. At a spectral resolution of R=100, the 5-11$\mu$m bandpass requires nearly twelve times as many transits ($\sim$94 transits) as the 3-11$\mu$m bandpass would require to obtain a 3.6$\sigma$ detection. Extending to 30$\mu$m incorporates the 17$\mu$m feature and only reduces the observational time to 72.8 transits while the constraints still exceed 1.50dex.

We find that the 3-5$\mu$m bandpass is sufficient to detect N$_{2}$O at 14.4 transits (R=100) or 23.8 transits (R=50), making it one of the only two molecular species that can be constrained in that bandpass within relatively few transits. However, that region is too narrow to provide adequate abundance constraints, resulting in a 1.27dex constraint at 50 transits. If we expand to either the 1-5$\mu$m or the 3-11$\mu$m bandpass, our required number of transits drops to 9.2 and 7.9, respectively. The corresponding constraints at 50 transits are 0.97dex (1-5$\mu$m) and 1.02dex (3-11$\mu$m), roughly an order-of-magnitude for both. The 3-11$\mu$m region would thus allow for a 3.6$\sigma$ detection in fewer transits than would be required for either O$_{3}$ or H$_{2}$O within the same regime. Likewise, the 1-5$\mu$m bandpass requires only 39.5\% of the observational time, to detect N$_{2}$O, as needed to detect H$_{2}$O in the same regime. With comparable abundance constraints, the 3-11$\mu$m bandpass proves to be the optimal bandpass, requiring only 74\% to 86\% (R=50 to R=100) of the observation time as the 1-5$\mu$m bandpass choice.

\section{Discussion} \label{Discussion}
Unsurprisingly, we find that obtaining the best detection significance values and abundance constraints for all of the molecules of interest are achieved by including the broadest possible wavelength coverage at the highest possible spectral resolution. Our primary goal, however, was to determine how the detection significance and abundance constraints behave as a function of the continuous wavelength coverage and spectral resolution. In the previous section, we elaborate on the effect that additional wavelength coverage and finer resolution have on the retrieval results for each molecule individually. Below we summarize our key findings:
\begin{itemize}
    \item When considering the contribution of additional wavelength coverage to the capability of an instrument with a continuous infrared bandpass, the most crucial spectral features (for H$_2$O, CO$_2$, CH$_4$, O$_3$, N$_2$O, and CO) in transmission spectra do not extend beyond 11$\mu$m.
    \item Utilizing a spectral resolution greater than R=100 proves to be significantly beneficial to resolving features at wavelengths shorter than 5$\mu$m. Redder wavelengths offer notably broader features, easily resolved with a resolution of R=50.
    \item Degeneracies caused by significantly overlapping spectral features are most easily resolved by including an additional feature for one or both molecules.
\end{itemize}
A brief summary of the crucial molecule-specific takeaways:
\begin{itemize}
    \item When attempting to detect and constrain CH$_{4}$, the observation of the 3.3$\mu$m feature is essential. If one must opt for lower resolutions (R=100), including the 2.3$\mu$m feature is incredibly valuable. While it is certainly ideal to include both features, it is crucial to acknowledge that the benefit of a higher resolution (R=300), in the absence of the 2.3$\mu$m feature, outweighs the benefit of including this feature at a lower resolution (R=100).
    \item The results of this study validate an alternative to the observing strategies which are currently available for H$_{2}$O, which are limited to the weak near-infrared features. We find that the more prominent unobscured spectral feature centered at 6.3$\mu$m, combined with the dense cluster of features from other molecules at 3-5$\mu$m, provides a significant decrease in the necessary observing time to detect H$_{2}$O compared to the 1-5$\mu$m bandpass. 
    \item CO$_{2}$, due to its very prominent features, is easily detectable even only observing the 4.3$\mu$m feature. However, increasing the wavelength coverage has statistically significant effects on the abundance constraints regardless of whether the broader bandpass includes additional CO$_{2}$ features.
    \item Given a m$_K=8.0$ source star, we find that utilizing an instrument with wavelength coverage from 3-11$\mu$m, a spectral resolution of only R=50, and a 25m$^2$ collecting area is sufficient to detect ozone abundances representative of modern-day Earth with less than 13 transits and to constrain the abundance value within an order-of-magnitude at 50 transits.
    \item All of the N$_{2}$O features from 1-30$\mu$m are partially or completely degenerate with features from CO$_{2}$ or CH$_{4}$. However, under the detector set-up evaluated in this study, observing only the features in the 3-5$\mu$m regime allows for a statistically significant detection in less than 15 transits. The 1-5$\mu$m and the 3-11$\mu$m bandpasses offer comparable improvements to both the detection significance and abundance constraints. 
\end{itemize}

The ability to detect and constrain the five biologically relevant molecular species in the atmosphere of an Earth-analog should be the benchmark by which the next-generation of exoplanet observatories are designed. The detection of significant amounts of CH$_{4}$ in tandem with O$_{3}$ and/or N$_{2}$O would constitute a promising indicator of biogenic origins. However, as it has been laboriously stated, CH$_{4}$ is the most challenging molecule to detect in this study and will prove to be the limiting variable in the development of a detector for these purposes. Unlike the other molecules considered here, CH$_{4}$ does not possess broad unobscured features in the mid-infrared and therefore does not have the advantage of relaxed spectral resolution requirements. It is important to acknowledge, however; that this is not an argument against observing mid-infrared wavelengths, which have been shown to be immensely valuable in the detection and constraint of all of the other molecular species in this study. Rather, our intention is to highlight that the choice of wavelength coverage for a next-generation near-to-mid-infrared spectrometer, particularly on the blue end of the bandpass, must be influenced by one or both of the near-infrared CH$_{4}$ features. To that end, we conclude that a near-to-mid-infrared spectrometer, aboard a next-generation space-based telescope (with a collecting area of 25m$^2$), boasting a bandpass including features between $\sim$2-11$\mu$m and a resolution of R$\simeq$50-300 would prove capable of detecting and constraining all of the key bio-indicator gasses in Earth's atmosphere. While the 2-11$\mu$m bandpass was not explicitly probed in this study, our retrieval results indicate that it would provide 3.6$\sigma$ detections in less transits and narrower abundance constraints than either the 1-5$\mu$m or 3-11$\mu$m bandpasses studies here.

\section{Conclusion} \label{Conclusion}
Our work has been motivated by the desire within the exoplanet community to characterize the atmospheres of temperate terrestrial planets beyond our Solar System. The detection of molecular biosignatures on these small rocky planets looms just beyond the capabilities of current space-based telescopes. A future facility with broad continuous wavelength coverage in the near and mid-infrared combined with detectors capable of driving observational noise down to the astrophysical noise floor will have what it takes to comprehensively probe the climate and composition of terrestrial exoplanet atmospheres. 

The 2020 Decadal Survey will consider four flagship mission concepts including one of particular interest to this study, the Origins Space Telescope. The Origins concept includes a near-to-mid-infrared spectrograph boasting an impressive 2.8-20$\mu$m continuous bandpass ranging from R$\simeq$50-300 with detectors designed to reduce instrument noise to approximately 5ppm (OST-STDT\footnote{\url{https://asd.gsfc.nasa.gov/firs/docs/OriginsVolume1MissionConceptStudyReport.pdf}}). Although it excludes the 2.3$\mu$m feature for CH$_{4}$, the design of this instrument strongly aligns with the findings in this work for the optimal observational setup for both detecting the presence and constraining the abundances of Earth-like biosignatures.

We look to expand upon this work by testing a subset of this analysis to a broader range of atmospheric compositions in order to determine the possibility of distinguishing several plausible atmospheric compositions on known terrestrial planets and those soon-to-be-discovered by \textit{TESS}. Additionally, we acknowledge the benefit of thermal emission spectroscopy for characterizing the thermal structure of exoplanetary atmospheres and providing additional information on the molecular abundances. Therefore, we aim to explore the effect of these observational parameters on a similar grid of synthetic thermal emission spectra. To this end, we anticipate that the synthesis of transmission and thermal emission as well as reflected light spectroscopy to be essential to the comprehensive understanding of these exoplanetary environments.

\software{{\tt PySynPhot} \citep{Lim15}, {\tt CHIMERA} \citep{Batalha17, Line13, Greene16,Line16}), {\tt PyMultiNest} \citep{Buchner14}}

\section*{Acknowledgements}
We would like to thank Tom Greene for providing us with his noise model. M.R. Line and L. Tremblay acknowledge the NASA XRP award NNX17AB56G for supporting this work. This work benefited from the 2018 Exoplanet Summer Program in the Other Worlds Laboratory (OWL) at the University of California, Santa Cruz, a program funded by the Heising-Simons Foundation. The authors also acknowledge NASA, which through the Origins Space Telescope mission concept study supported travel to the Other Worlds Laboratory Summer Program to facilitate this work.  

Part of the research was carried out at the Jet Propulsion Laboratory, California Institute of Technology, under contract with the National Aeronautics and Space Administration. 

\bibliography{refs}

\begin{thebibliography}{}
\expandafter\ifx\csname natexlab\endcsname\relax\def\natexlab#1{#1}\fi
\providecommand{\url}[1]{\href{#1}{#1}}

\bibitem[{{Barclay} {et~al.}(2018){Barclay}, {Pepper}, \&
  {Quintana}}]{Barcklay18}
{Barclay}, T., {Pepper}, J., \& {Quintana}, E.~V. 2018, \apjs, 239, 2

\bibitem[{{Barstow} {et~al.}(2014){Barstow}, {Aigrain}, {Irwin}, {Hackler},
  {Fletcher}, {Lee}, \& {Gibson}}]{Barstow14}
{Barstow}, J.~K., {Aigrain}, S., {Irwin}, P.~G.~J., {et~al.} 2014, \apj, 786,
  154

\bibitem[{{Barstow} \& {Irwin}(2016)}]{Barstow16}
{Barstow}, J.~K., \& {Irwin}, P.~G.~J. 2016, \mnras, 461, L92

\bibitem[{{Batalha} {et~al.}(2018){Batalha}, {Lewis}, {Line}, {Valenti}, \&
  {Stevenson}}]{Batalha18}
{Batalha}, N.~E., {Lewis}, N.~K., {Line}, M.~R., {Valenti}, J., \& {Stevenson},
  K. 2018, \apjl, 856, L34

\bibitem[{{Batalha} \& {Line}(2017)}]{Batalha17}
{Batalha}, N.~E., \& {Line}, M.~R. 2017, \aj, 153, 151

\bibitem[{{Beichman} {et~al.}(2014){Beichman}, {Benneke}, {Knutson}, {Smith},
  {Lagage}, {Dressing}, {Latham}, {Lunine}, {Birkmann}, {Ferruit}, {Giardino},
  {Kempton}, {Carey}, {Krick}, {Deroo}, {Mand ell}, {Ressler}, {Shporer},
  {Swain}, {Vasisht}, {Ricker}, {Bouwman}, {Crossfield}, {Greene}, {Howell},
  {Christiansen}, {Ciardi}, {Clampin}, {Greenhouse}, {Sozzetti}, {Goudfrooij},
  {Hines}, {Keyes}, {Lee}, {McCullough}, {Robberto}, {Stansberry}, {Valenti},
  {Rieke}, {Rieke}, {Fortney}, {Bean}, {Kreidberg}, {Ehrenreich}, {Deming},
  {Albert}, {Doyon}, \& {Sing}}]{Beichman14}
{Beichman}, C., {Benneke}, B., {Knutson}, H., {et~al.} 2014, \pasp, 126, 1134

\bibitem[{{Benneke} \& {Seager}(2012)}]{Benneke12}
{Benneke}, B., \& {Seager}, S. 2012, \apj, 753, 100

\bibitem[{{Benneke} \& {Seager}(2013)}]{Benneke13}
---. 2013, \apj, 778, 153

\bibitem[{{Buchner} {et~al.}(2014){Buchner}, {Georgakakis}, {Nandra}, {Hsu},
  {Rangel}, {Brightman}, {Merloni}, {Salvato}, {Donley}, \&
  {Kocevski}}]{Buchner14}
{Buchner}, J., {Georgakakis}, A., {Nandra}, K., {et~al.} 2014, \aap, 564, A125

\bibitem[{Catling \& Kasting(2007)}]{CatKast07}
Catling, D., \& Kasting, J. 2007, Planetary Atmospheres and Life, ed.
  W.~Sullivan \& J.~Baross (United Kingdom: Cambridge University Press), 91 --
  116

\bibitem[{{Chapman} {et~al.}(2017){Chapman}, {Zellem}, {Line}, {Vasisht},
  {Bryden}, {Willacy}, {Iyer}, {Bean}, {Cowan}, {Fortney}, {Griffith},
  {Kataria}, {Kempton}, {Kreidberg}, {Moses}, {Stevenson}, \&
  {Swain}}]{Chapman17}
{Chapman}, J.~W., {Zellem}, R.~T., {Line}, M.~R., {et~al.} 2017, \pasp, 129,
  104402

\bibitem[{{Cockell} {et~al.}(2009){Cockell}, {Kaltenegger}, \&
  {Raven}}]{Cockell09}
{Cockell}, C.~S., {Kaltenegger}, L., \& {Raven}, J.~A. 2009, Astrobiology, 9,
  623

\bibitem[{{Cowan} {et~al.}(2015){Cowan}, {Greene}, {Angerhausen}, {Batalha},
  {Clampin}, {Col{\'o}n}, {Crossfield}, {Fortney}, {Gaudi}, {Harrington},
  {Iro}, {Lillie}, {Linsky}, {Lopez-Morales}, {Mandell}, \&
  {Stevenson}}]{Cowan15}
{Cowan}, N.~B., {Greene}, T., {Angerhausen}, D., {et~al.} 2015, \pasp, 127, 311

\bibitem[{{Deming} {et~al.}(2009){Deming}, {Seager}, {Winn}, {Miller-Ricci},
  {Clampin}, {Lindler}, {Greene}, {Charbonneau}, {Laughlin}, {Ricker},
  {Latham}, \& {Ennico}}]{Deming09}
{Deming}, D., {Seager}, S., {Winn}, J., {et~al.} 2009, \pasp, 121, 952

\bibitem[{{Domagal-Goldman} {et~al.}(2014){Domagal-Goldman}, {Segura},
  {Claire}, {Robinson}, \& {Meadows}}]{DomGold14}
{Domagal-Goldman}, S.~D., {Segura}, A., {Claire}, M.~W., {Robinson}, T.~D., \&
  {Meadows}, V.~S. 2014, \apj, 792, 90

\bibitem[{{Gillon} {et~al.}(2016){Gillon}, {Jehin}, {Lederer}, {Delrez}, {de
  Wit}, {Burdanov}, {Van Grootel}, {Burgasser}, {Triaud}, {Opitom}, {Demory},
  {Sahu}, {Bardalez Gagliuffi}, {Magain}, \& {Queloz}}]{Gillon16}
{Gillon}, M., {Jehin}, E., {Lederer}, S.~M., {et~al.} 2016, \nat, 533, 221

\bibitem[{{Gillon} {et~al.}(2017){Gillon}, {Triaud}, {Demory}, {Jehin}, {Agol},
  {Deck}, {Lederer}, {de Wit}, {Burdanov}, {Ingalls}, {Bolmont}, {Leconte},
  {Raymond}, {Selsis}, {Turbet}, {Barkaoui}, {Burgasser}, {Burleigh}, {Carey},
  {Chaushev}, {Copperwheat}, {Delrez}, {Fernand es}, {Holdsworth}, {Kotze},
  {Van Grootel}, {Almleaky}, {Benkhaldoun}, {Magain}, \& {Queloz}}]{Gillon17}
{Gillon}, M., {Triaud}, A. H.~M.~J., {Demory}, B.-O., {et~al.} 2017, \nat, 542,
  456

\bibitem[{{Glass} \& {Blair}(2015)}]{Glass15a}
{Glass}, J.~D., \& {Blair}, W.~D. 2015, IEEE Transactions on Aerospace
  Electronic Systems, 51, 1927

\bibitem[{{Glass} {et~al.}(2015){Glass}, {Blair}, \& {Lanterman}}]{Glass15b}
{Glass}, J.~D., {Blair}, W.~D., \& {Lanterman}, A.~D. 2015, IEEE Transactions
  on Signal Processing, 63, 6673

\bibitem[{Gordon {et~al.}(2017)Gordon, Rothman, Hill, Kochanov, Tan, Bernath,
  Birk, Boudon, Campargue, Chance, Drouin, Flaud, Gamache, Hodges, Jacquemart,
  Perevalov, Perrin, Shine, Smith, Tennyson, Toon, Tran, Tyuterev, Barbe,
  Császár, Devi, Furtenbacher, Harrison, Hartmann, Jolly, Johnson, Karman,
  Kleiner, Kyuberis, Loos, Lyulin, Massie, Mikhailenko, Moazzen-Ahmadi,
  Müller, Naumenko, Nikitin, Polyansky, Rey, Rotger, Sharpe, Sung, Starikova,
  Tashkun, Auwera, Wagner, Wilzewski, Wcisło, Yu, \& Zak}]{Gordon17}
Gordon, I., Rothman, L., Hill, C., {et~al.} 2017, Journal of Quantitative
  Spectroscopy and Radiative Transfer, 203, 3

\bibitem[{{Greene} {et~al.}(2016){Greene}, {Line}, {Montero}, {Fortney},
  {Lustig-Yaeger}, \& {Luther}}]{Greene16}
{Greene}, T.~P., {Line}, M.~R., {Montero}, C., {et~al.} 2016, \apj, 817, 17

\bibitem[{{Grimm} {et~al.}(2018){Grimm}, {Demory}, {Gillon}, {Dorn}, {Agol},
  {Burdanov}, {Delrez}, {Sestovic}, {Triaud}, {Turbet}, {Bolmont}, {Caldas},
  {de Wit}, {Jehin}, {Leconte}, {Raymond}, {Van Grootel}, {Burgasser}, {Carey},
  {Fabrycky}, {Heng}, {Hernandez}, {Ingalls}, {Lederer}, {Selsis}, \&
  {Queloz}}]{Grimm18}
{Grimm}, S.~L., {Demory}, B.-O., {Gillon}, M., {et~al.} 2018, \aap, 613, A68

\bibitem[{{Hardegree-Ullman} {et~al.}(2019){Hardegree-Ullman}, {Cushing},
  {Muirhead}, \& {Christiansen}}]{HardUll19}
{Hardegree-Ullman}, K.~K., {Cushing}, M.~C., {Muirhead}, P.~S., \&
  {Christiansen}, J.~L. 2019, \aj, 158, 75

\bibitem[{{Harman} {et~al.}(2018){Harman}, {Felton}, {Hu}, {Domagal-Goldman},
  {Segura}, {Tian}, \& {Kasting}}]{Harman18}
{Harman}, C.~E., {Felton}, R., {Hu}, R., {et~al.} 2018, \apj, 866, 56

\bibitem[{{Harman} {et~al.}(2015){Harman}, {Schwieterman}, {Schottelkotte}, \&
  {Kasting}}]{Harman15}
{Harman}, C.~E., {Schwieterman}, E.~W., {Schottelkotte}, J.~C., \& {Kasting},
  J.~F. 2015, \apj, 812, 137

\bibitem[{{Husser} {et~al.}(2013){Husser}, {Wende-von Berg}, {Dreizler},
  {Homeier}, {Reiners}, {Barman}, \& {Hauschildt}}]{husser2013}
{Husser}, T.~O., {Wende-von Berg}, S., {Dreizler}, S., {et~al.} 2013, \aap,
  553, A6

\bibitem[{{Kasting}(2009)}]{Kasting09}
{Kasting}, J.~F. 2009, Geochimica et Cosmochimica Acta Supplement, 73, A625

\bibitem[{{Kempton} {et~al.}(2018){Kempton}, {Bean}, {Louie}, {Deming}, {Koll},
  {Mansfield}, {Christiansen}, {L{\'o}pez-Morales}, {Swain}, {Zellem},
  {Ballard}, {Barclay}, {Barstow}, {Batalha}, {Beatty}, {Berta-Thompson},
  {Birkby}, {Buchhave}, {Charbonneau}, {Cowan}, {Crossfield}, {de Val-Borro},
  {Doyon}, {Dragomir}, {Gaidos}, {Heng}, {Hu}, {Kane}, {Kreidberg}, {Mallonn},
  {Morley}, {Narita}, {Nascimbeni}, {Pall{\'e}}, {Quintana}, {Rauscher},
  {Seager}, {Shkolnik}, {Sing}, {Sozzetti}, {Stassun}, {Valenti}, \& {von
  Essen}}]{Kempton18}
{Kempton}, E. M.~R., {Bean}, J.~L., {Louie}, D.~R., {et~al.} 2018, \pasp, 130,
  114401

\bibitem[{Kochanov {et~al.}(2016)Kochanov, Gordon, Rothman, Wcisło, Hill, \&
  Wilzewski}]{Kochonov16}
Kochanov, R., Gordon, I., Rothman, L., {et~al.} 2016, Journal of Quantitative
  Spectroscopy and Radiative Transfer, 177, 15 , xVIIIth Symposium on High
  Resolution Molecular Spectroscopy (HighRus-2015), Tomsk, Russia

\bibitem[{{Krissansen-Totton} {et~al.}(2016){Krissansen-Totton}, {Bergsman}, \&
  {Catling}}]{krissTott16}
{Krissansen-Totton}, J., {Bergsman}, D.~S., \& {Catling}, D.~C. 2016,
  Astrobiology, 16, 39

\bibitem[{{Krissansen-Totton} {et~al.}(2018){Krissansen-Totton}, {Garland},
  {Irwin}, \& {Catling}}]{KT18}
{Krissansen-Totton}, J., {Garland}, R., {Irwin}, P., \& {Catling}, D.~C. 2018,
  \aj, 156, 114

\bibitem[{{L{\'e}ger}(2000)}]{Leger00}
{L{\'e}ger}, A. 2000, Advances in Space Research, 25, 2209

\bibitem[{{L{\'e}ger} {et~al.}(1993){L{\'e}ger}, {Pirre}, \&
  {Marceau}}]{Leger93}
{L{\'e}ger}, A., {Pirre}, M., \& {Marceau}, F.~J. 1993, \aap, 277, 309

\bibitem[{Lim {et~al.}(2015)Lim, Diaz, \& Laidler}]{Lim15}
Lim, P., Diaz, R., \& Laidler, V. 2015, Astrophysics Source Code Library

\bibitem[{{Line} \& {Parmentier}(2016)}]{Line16}
{Line}, M.~R., \& {Parmentier}, V. 2016, \apj, 820, 78

\bibitem[{{Line} {et~al.}(2012){Line}, {Zhang}, {Vasisht}, {Natraj}, {Chen}, \&
  {Yung}}]{Line12}
{Line}, M.~R., {Zhang}, X., {Vasisht}, G., {et~al.} 2012, \apj, 749, 93

\bibitem[{{Line} {et~al.}(2013){Line}, {Wolf}, {Zhang}, {Knutson}, {Kammer},
  {Ellison}, {Deroo}, {Crisp}, \& {Yung}}]{Line13}
{Line}, M.~R., {Wolf}, A.~S., {Zhang}, X., {et~al.} 2013, \apj, 775, 137

\bibitem[{{Lustig-Yaeger} {et~al.}(2019){Lustig-Yaeger}, {Meadows}, \&
  {Lincowski}}]{LustYeag19}
{Lustig-Yaeger}, J., {Meadows}, V.~S., \& {Lincowski}, A.~P. 2019, \aj, 158, 27

\bibitem[{Matsuo {et~al.}(2018)Matsuo, Greene, Roellig, McMurray, Johnson,
  Kashani, Goda, Ido, Ito, Tsuboi, Yamamuro, Ikeda, Shibai, Sumi, Sakon, \&
  Ennico-Smith}]{10.1117/12.2311896}
Matsuo, T., Greene, T., Roellig, T.~L., {et~al.} 2018, in Space Telescopes and
  Instrumentation 2018: Optical, Infrared, and Millimeter Wave, Vol. 10698,
  International Society for Optics and Photonics (SPIE), 1218 -- 1229

\bibitem[{{Meadows}(2017)}]{Meadows17}
{Meadows}, V.~S. 2017, Astrobiology, 17, 1022

\bibitem[{{Meadows} {et~al.}(2018){Meadows}, {Reinhard}, {Arney}, {Parenteau},
  {Schwieterman}, {Domagal-Goldman}, {Lincowski}, {Stapelfeldt}, {Rauer},
  {DasSarma}, {Hegde}, {Narita}, {Deitrick}, {Lustig-Yaeger}, {Lyons},
  {Siegler}, \& {Grenfell}}]{Meadows18}
{Meadows}, V.~S., {Reinhard}, C.~T., {Arney}, G.~N., {et~al.} 2018,
  Astrobiology, 18, 630

\bibitem[{{Morley} {et~al.}(2017){Morley}, {Kreidberg}, {Rustamkulov},
  {Robinson}, \& {Fortney}}]{Morley17}
{Morley}, C.~V., {Kreidberg}, L., {Rustamkulov}, Z., {Robinson}, T., \&
  {Fortney}, J.~J. 2017, \apj, 850, 121

\bibitem[{{Quirrenbach} {et~al.}(2018){Quirrenbach}, {Amado}, {Ribas},
  {Reiners}, {Caballero}, {Seifert}, {Aceituno}, {Azzaro}, {Baroch}, {Barrado},
  {Bauer}, {Becerril}, {B{\`e}jar}, {Ben{\'\i}tez}, {Brinkm{\"o}ller}, {Cardona
  Guill{\'e}n}, {Cifuentes}, {Colom{\'e}}, {Cort{\'e}s-Contreras}, {Czesla},
  {Dreizler}, {Fr{\"o}lich}, {Fuhrmeister}, {Galad{\'\i}-Enr{\'\i}quez},
  {Gonz{\'a}lez Hern{\'a}ndez}, {Gonz{\'a}lez Peinado}, {Guenther}, {de
  Guindos}, {Hagen}, {Hatzes}, {Hauschildt}, {Helmling}, {Henning}, {Herbort},
  {Hern{\'a}ndez Casta{\~n}o}, {Herrero}, {Hintz}, {Jeffers}, {Johnson}, {de
  Juan}, {Kaminski}, {Klahr}, {K{\"u}rster}, {Lafarga}, {Sairam}, {Lamp{\'o}n},
  {Lara}, {Launhardt}, {L{\'o}pez del Fresno}, {L{\'o}pez-Puertas}, {Luque},
  {Mandel}, {Marfil}, {Mart{\'\i}n}, {Mart{\'\i}n-Ruiz}, {Mathar}, {Montes},
  {Morales}, {Nagel}, {Nortmann}, {Nowak}, {Pall{\'e}}, {Passegger}, {Pavlov},
  {Pedraz}, {P{\'e}rez-Medialdea}, {Perger}, {Rebolo}, {Reffert},
  {Rodr{\'\i}guez}, {Rodr{\'\i}guez L{\'o}pez}, {Rosich}, {Sabotta}, {Sadegi},
  {Salz}, {S{\'a}nchez-L{\'o}pez}, {Sanz-Forcada}, {Sarkis}, {Sch{\"a}fer},
  {Schiller}, {Schmitt}, {Sch{\"o}fer}, {Schweitzer}, {Shulyak}, {Solano},
  {Stahl}, {Tala Pinto}, {Trifonov}, {Zapatero Osorio}, {Yan}, {Zechmeister},
  {Abell{\'a}n}, {Abril}, {Alonso-Floriano}, {Ammler-von Eiff},
  {Anglada-Escud{\'e}}, {Anwand-Heerwart}, {Arroyo-Torres}, {Berdi{\~n}as},
  {Bergondy}, {Bl{\"u}mcke}, {del Burgo}, {Cano}, {Carro}, {C{\'a}rdenas},
  {Casal}, {Claret}, {D{\'\i}ez-Alonso}, {Doellinger}, {Dorda}, {Feiz},
  {Fern{\'a}ndez}, {Ferro}, {Gaisn{\'e}}, {Gallardo}, {G{\'a}lvez-Ortiz},
  {Garc{\'\i}a-Piquer}, {Garc{\'\i}a-Vargas}, {Garrido}, {Gesa}, {G{\'o}mez
  Galera}, {Gonz{\'a}lez-{\'A}lvarez}, {Gonz{\'a}lez-Cuesta}, {Grohnert},
  {Gr{\"o}zinger}, {Gu{\`a}rdia}, {Guijarro}, {Hedrosa}, {Hermann}, {Hermelo},
  {Hern{\'a}ndez Arab{\'\i}}, {Hern{\'a}ndez Hernando}, {Hidalgo}, {Holgado},
  {Huber}, {Huber}, {Huke}, {Kehr}, {Kim}, {Klein}, {Kl{\"u}ter}, {Klutsch},
  {Labarga}, {Labiche}, {Lamert}, {Laun}, {L{\'a}zaro}, {Lemke}, {Lenzen},
  {Llamas}, {Lizon}, {Lodieu}, {L{\'o}pez Gonz{\'a}lez}, {L{\'o}pez-Morales},
  {L{\'o}pez Salas}, {L{\'o}pez-Santiago}, {Mag{\'a}n Madinabeitia}, {Mall},
  {Mancini}, {Mar{\'\i}n Molina}, {Mart{\'\i}nez-Rodr{\'\i}guez}, {Maroto
  Fern{\'a}ndez}, {Marvin}, {Mirabet}, {Moreno-Raya}, {Moya}, {Mundt},
  {Naranjo}, {Panduro}, {Pascual}, {P{\'e}rez-Calpena}, {Perryman}, {Pluto},
  {Ram{\'o}n}, {Redondo}, {Reinhart}, {Rhode}, {Rix}, {Rodler}, {Rohloff},
  {S{\'a}nchez-Blanco}, {S{\'a}nchez Carrasco}, {Sarmiento}, {Schmidt},
  {Storz}, {Strachan}, {St{\"u}rmer}, {Su{\'a}rez}, {Tabernero}, {Tal-Or},
  {Tulloch}, {Ulbrich}, {Veredas}, {Vico Linares}, {Vidal-Dasilva},
  {Vilardell}, {Wagner}, {Winkler}, {Wolthoff}, {Xu}, \&
  {Zhao}}]{Quirrenbach18}
{Quirrenbach}, A., {Amado}, P.~J., {Ribas}, I., {et~al.} 2018, in Society of
  Photo-Optical Instrumentation Engineers (SPIE) Conference Series, Vol. 10702,
  \procspie, 107020W

\bibitem[{{Robinson} {et~al.}(2017){Robinson}, {Fortney}, \&
  {Hubbard}}]{Robinson17}
{Robinson}, T.~D., {Fortney}, J.~J., \& {Hubbard}, W.~B. 2017, \apj, 850, 128

\bibitem[{{Rocchetto} {et~al.}(2016){Rocchetto}, {Waldmann}, {Venot}, {Lagage},
  \& {Tinetti}}]{Rochetto16}
{Rocchetto}, M., {Waldmann}, I.~P., {Venot}, O., {Lagage}, P.~O., \& {Tinetti},
  G. 2016, \apj, 833, 120

\bibitem[{{Sagan} {et~al.}(1993){Sagan}, {Thompson}, {Carlson}, {Gurnett}, \&
  {Hord}}]{Sagan93}
{Sagan}, C., {Thompson}, W.~R., {Carlson}, R., {Gurnett}, D., \& {Hord}, C.
  1993, \nat, 365, 715

\bibitem[{{Schwieterman} {et~al.}(2015){Schwieterman}, {Robinson}, {Meadows},
  {Misra}, \& {Domagal-Goldman}}]{Schwieterman15}
{Schwieterman}, E.~W., {Robinson}, T.~D., {Meadows}, V.~S., {Misra}, A., \&
  {Domagal-Goldman}, S. 2015, \apj, 810, 57

\bibitem[{{Schwieterman} {et~al.}(2016){Schwieterman}, {Meadows},
  {Domagal-Goldman}, {Deming}, {Arney}, {Luger}, {Harman}, {Misra}, \&
  {Barnes}}]{Schwiterman16}
{Schwieterman}, E.~W., {Meadows}, V.~S., {Domagal-Goldman}, S.~D., {et~al.}
  2016, \apjl, 819, L13

\bibitem[{{Seager}(2014)}]{Seager14}
{Seager}, S. 2014, Proceedings of the National Academy of Science, 111, 12634

\bibitem[{{Seager} \& {Bains}(2015)}]{Seager15}
{Seager}, S., \& {Bains}, W. 2015, Science Advances, 1, e1500047

\bibitem[{{Seager} \& {Deming}(2010)}]{Seager10}
{Seager}, S., \& {Deming}, D. 2010, \araa, 48, 631

\bibitem[{{Segura} {et~al.}(2003){Segura}, {Krelove}, {Kasting}, {Sommerlatt},
  {Meadows}, {Crisp}, {Cohen}, \& {Mlawer}}]{Segura03}
{Segura}, A., {Krelove}, K., {Kasting}, J.~F., {et~al.} 2003, Astrobiology, 3,
  689

\bibitem[{{Stevenson} {et~al.}(2016){Stevenson}, {Lewis}, {Bean}, {Beichman},
  {Fraine}, {Kilpatrick}, {Krick}, {Lothringer}, {Mandell}, {Valenti}, {Agol},
  {Angerhausen}, {Barstow}, {Birkmann}, {Burrows}, {Charbonneau}, {Cowan},
  {Crouzet}, {Cubillos}, {Curry}, {Dalba}, {de Wit}, {Deming}, {D{\'e}sert},
  {Doyon}, {Dragomir}, {Ehrenreich}, {Fortney}, {Garc{\'\i}a Mu{\~n}oz},
  {Gibson}, {Gizis}, {Greene}, {Harrington}, {Heng}, {Kataria}, {Kempton},
  {Knutson}, {Kreidberg}, {Lafreni{\`e}re}, {Lagage}, {Line}, {Lopez-Morales},
  {Madhusudhan}, {Morley}, {Rocchetto}, {Schlawin}, {Shkolnik}, {Shporer},
  {Sing}, {Todorov}, {Tucker}, \& {Wakeford}}]{Stevenson16}
{Stevenson}, K.~B., {Lewis}, N.~K., {Bean}, J.~L., {et~al.} 2016, \pasp, 128,
  094401

\bibitem[{{Suissa} {et~al.}(2018){Suissa}, {Chen}, \& {Kipping}}]{Suissa18a}
{Suissa}, G., {Chen}, J., \& {Kipping}, D. 2018, \mnras, 476, 2613

\bibitem[{{Suissa} \& {Kipping}(2018)}]{Suissa18b}
{Suissa}, G., \& {Kipping}, D. 2018, Research Notes of the American
  Astronomical Society, 2, 31

\bibitem[{{Tian} {et~al.}(2014){Tian}, {France}, {Linsky}, {Mauas}, \&
  {Vieytes}}]{Tian14}
{Tian}, F., {France}, K., {Linsky}, J.~L., {Mauas}, P. J.~D., \& {Vieytes},
  M.~C. 2014, Earth and Planetary Science Letters, 385, 22

\bibitem[{{Trotta}(2008)}]{Trotta08}
{Trotta}, R. 2008, Contemporary Physics, 49, 71

\bibitem[{{Turbet} {et~al.}(2018){Turbet}, {Bolmont}, {Leconte}, {Forget},
  {Selsis}, {Tobie}, {Caldas}, {Naar}, \& {Gillon}}]{Turbet18}
{Turbet}, M., {Bolmont}, E., {Leconte}, J., {et~al.} 2018, \aap, 612, A86

\bibitem[{{Unterborn} \& {Panero}(2017)}]{Unterborn17}
{Unterborn}, C.~T., \& {Panero}, W.~R. 2017, \apj, 845, 61

\bibitem[{{Wolf}(2017)}]{Wolf17}
{Wolf}, E.~T. 2017, \apjl, 839, L1

\bibitem[{{Wunderlich} {et~al.}(2019){Wunderlich}, {Godolt}, {Grenfell},
  {St{\"a}dt}, {Smith}, {Gebauer}, {Schreier}, {Hedelt}, \&
  {Rauer}}]{Wunderlich19}
{Wunderlich}, F., {Godolt}, M., {Grenfell}, J.~L., {et~al.} 2019, \aap, 624,
  A49

\bibitem[{{Zechmeister} {et~al.}(2019){Zechmeister}, {Dreizler}, {Ribas},
  {Reiners}, {Caballero}, {Bauer}, {B{\'e}jar}, {Gonz{\'a}lez-Cuesta},
  {Herrero}, {Lalitha}, {L{\'o}pez-Gonz{\'a}lez}, {Luque}, {Morales},
  {Pall{\'e}}, {Rodr{\'\i}guez}, {Rodr{\'\i}guez L{\'o}pez}, {Tal-Or},
  {Anglada-Escud{\'e}}, {Quirrenbach}, {Amado}, {Abril}, {Aceituno},
  {Aceituno}, {Alonso-Floriano}, {Ammler-von Eiff}, {Antona Jim{\'e}nez},
  {Anwand-Heerwart}, {Arroyo-Torres}, {Azzaro}, {Baroch}, {Barrado},
  {Becerril}, {Ben{\'\i}tez}, {Berdi{\~n}as}, {Bergond}, {Bluhm},
  {Brinkm{\"o}ller}, {del Burgo}, {Calvo Ortega}, {Cano}, {Cardona
  Guill{\'e}n}, {Carro}, {C{\'a}rdenas V{\'a}zquez}, {Casal},
  {Casasayas-Barris}, {Casanova}, {Chaturvedi}, {Cifuentes}, {Claret},
  {Colom{\'e}}, {Cort{\'e}s-Contreras}, {Czesla}, {D{\'\i}ez-Alonso}, {Dorda},
  {Fern{\'a}ndez}, {Fern{\'a}ndez-Mart{\'\i}n}, {Fuhrmeister}, {Fukui},
  {Galad{\'\i}-Enr{\'\i}quez}, {Gallardo Cava}, {Garcia de la Fuente},
  {Garcia-Piquer}, {Garc{\'\i}a Vargas}, {Gesa}, {G{\'o}ngora Rueda},
  {Gonz{\'a}lez-{\'A}lvarez}, {Gonz{\'a}lez Hern{\'a}ndez},
  {Gonz{\'a}lez-Peinado}, {Gr{\"o}zinger}, {Gu{\`a}rdia}, {Guijarro}, {de
  Guindos}, {Hatzes}, {Hauschildt}, {Hedrosa}, {Helmling}, {Henning},
  {Hermelo}, {Hern{\'a}ndez Arabi}, {Hern{\'a}ndez Casta{\~n}o}, {Hern{\'a}ndez
  Otero}, {Hintz}, {Huke}, {Huber}, {Jeffers}, {Johnson}, {de Juan},
  {Kaminski}, {Kemmer}, {Kim}, {Klahr}, {Klein}, {Kl{\"u}ter}, {Klutsch},
  {Kossakowski}, {K{\"u}rster}, {Labarga}, {Lafarga}, {Llamas}, {Lamp{\'o}n},
  {Lara}, {Launhardt}, {L{\'a}zaro}, {Lodieu}, {L{\'o}pez del Fresno},
  {L{\'o}pez-Puertas}, {L{\'o}pez Salas}, {L{\'o}pez-Santiago}, {Mag{\'a}n
  Madinabeitia}, {Mall}, {Mancini}, {Mand el}, {Marfil}, {Mar{\'\i}n Molina},
  {Maroto Fern{\'a}ndez}, {Mart{\'\i}n}, {Mart{\'\i}n-Fern{\'a}ndez},
  {Mart{\'\i}n-Ruiz}, {Marvin}, {Mirabet}, {Monta{\~n}{\'e}s-Rodr{\'\i}guez},
  {Montes}, {Moreno-Raya}, {Nagel}, {Naranjo}, {Narita}, {Nortmann}, {Nowak},
  {Ofir}, {Oshagh}, {Panduro}, {Parviainen}, {Pascual}, {Passegger}, {Pavlov},
  {Pedraz}, {P{\'e}rez-Calpena}, {P{\'e}rez Medialdea}, {Perger}, {Perryman},
  {Rabaza}, {Ram{\'o}n Ballesta}, {Rebolo}, {Redondo}, {Reffert}, {Reinhardt},
  {Rhode}, {Rix}, {Rodler}, {Rodr{\'\i}guez Trinidad}, {Rosich}, {Sadegi},
  {S{\'a}nchez-Blanco}, {S{\'a}nchez Carrasco}, {S{\'a}nchez-L{\'o}pez},
  {Sanz-Forcada}, {Sarkis}, {Sarmiento}, {Sch{\"a}fer}, {Schmitt},
  {Sch{\"o}fer}, {Schweitzer}, {Seifert}, {Shulyak}, {Solano}, {Sota}, {Stahl},
  {Stock}, {Strachan}, {Stuber}, {St{\"u}rmer}, {Su{\'a}rez}, {Tabernero},
  {Tala Pinto}, {Trifonov}, {Veredas}, {Vico Linares}, {Vilardell}, {Wagner},
  {Wolthoff}, {Xu}, {Yan}, \& {Zapatero Osorio}}]{Zechmeister19}
{Zechmeister}, M., {Dreizler}, S., {Ribas}, I., {et~al.} 2019, \aap, 627, A49

\end{thebibliography}
\nocite{*}

\appendixpage
\begin{figure}[ht]
    \centering
    \includegraphics[width=\columnwidth]{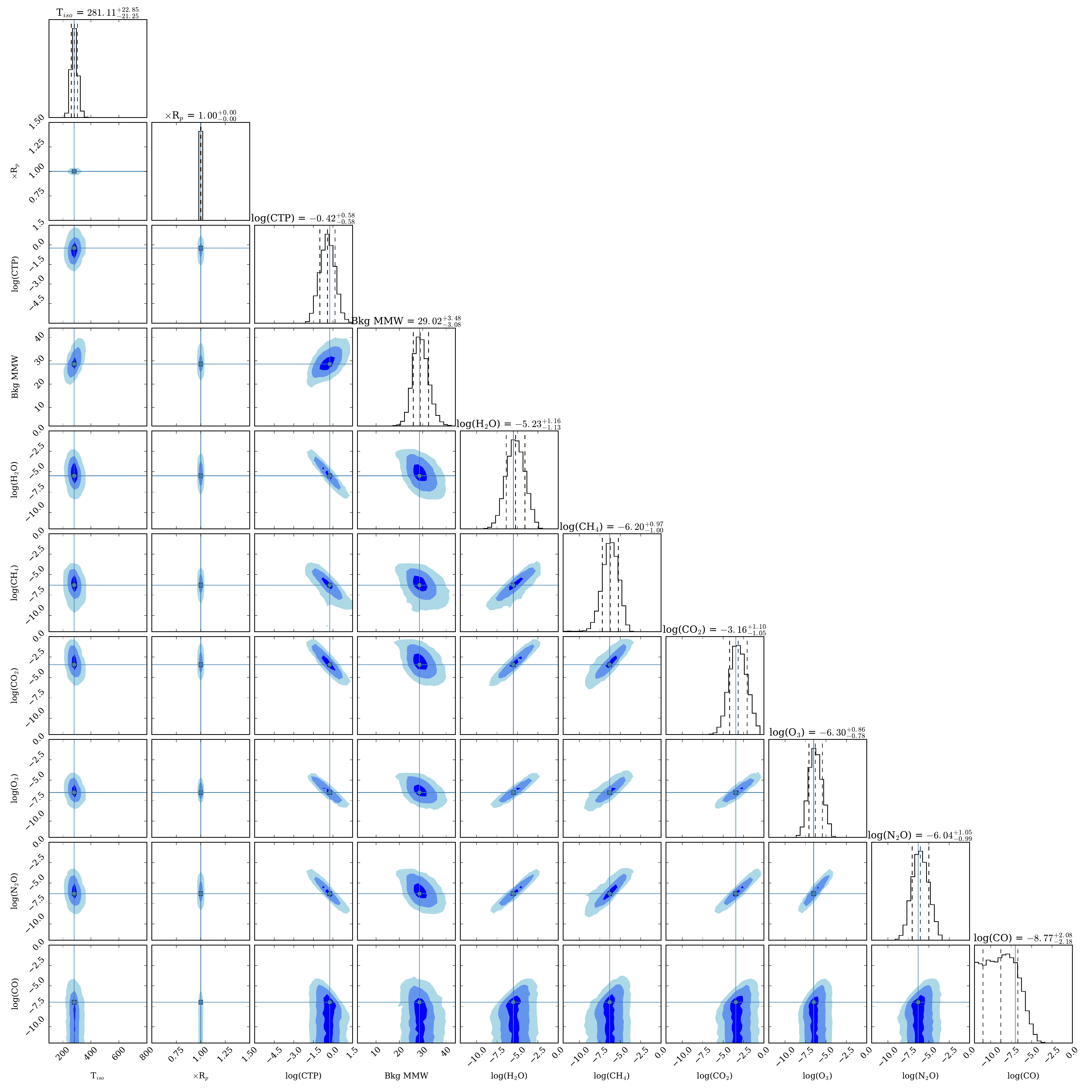}
    \caption{Corner plot for the 3-11$\mu$m at R=100 test case. Shows the posterior probability distributions for all of the parameters in our model.  In the dropbox link (presented earlier) we provide all of the corner plots for the R=50 and R=100 scenarios.}
\end{figure}

\end{document}